\documentclass{article}
\usepackage[utf8]{inputenc}
\usepackage{amsmath}
\usepackage{physics}
\usepackage{subcaption}
\usepackage{graphicx}
\usepackage{xcolor}
\usepackage{amssymb}
\usepackage[sort&compress]{natbib}
\usepackage{braket}
\usepackage[affil-it]{authblk}
\usepackage{vmargin}
\usepackage{multicol}
\usepackage[colorlinks=true,citecolor=red,urlcolor=blue,linkcolor=blue]{hyperref}

\title{Vector Dark Matter in the Fundamental Representation of $SU(2)_{L}$: Sommerfeld Enhancement and Indirect Detection}
\author[a,b,c]{Sebastián Acevedo Espinoza\thanks{sebastian.acevedoe@sansano.usm.cl}}
\author[a]{Amanda Rodríguez \thanks{amanda.rodriguez@usm.cl}}
\author[a,b,c]{Alfonso Zerwekh \thanks{alfonso.zerwekh@usm.cl}}
\affil[a]{Departamento de Física, Universidad Técnica Federico santa María, Avenida España 1680, Valparaı\'iso, Chile}
\affil[b]{Millennium Institute for Subatomic Physics at High Energy Frontier – SAPHIR,
Fernandez Concha 700, Santiago, Chile}
\affil[c]{Centro Cient\'ifico-Tecnol\'ogico de Valpara\'iso, Universidad T\'ecnica Federico Santa Mar\'ia,
Avenida España 1680, Valpara\'iso, Chile}

\begin{document}
\maketitle
\begin{abstract}

In this work, we study an extension of the Standard Model that includes a new massive vector field in the fundamental representation of $SU(2)_{L}$. The neutral component of this field provides a natural dark matter candidate. We compute the annihilation cross section including Sommerfeld enhancement and the gamma-ray flux arising from dark matter annihilation. We derive constraints on the model parameter space using current gamma-ray observations and investigate the prospects for future searches. We find that the model exhibits resonances for dark matter masses in the range $2-10$ TeV, whose properties are influenced by the value of the Higgs portal coupling. We show that part of the remaining parameter space can be probed by CTA in the near future.  

\end{abstract}

\maketitle

\vspace{1.0cm}
\newpage
\section{Introduction}
Currently, one of the most urgent problems in particle physics revolves around the elucidation of Dark Matter's (DM) fundamental nature ~\cite{Bertone:2004pz,Arkani-Hamed:2008hhe}. It is well known that, to account for the overarching phenomenology of DM, it suffices to posit the existence of a new particle falling within a mass range spanning from a few GeV to a few TeV, characterized by an interaction strength with the standard sector on the order of the weak interaction. This phenomenon is commonly referred to as the WIMP miracle, where WIMP stands for Weakly Interacting Massive Particle~\cite{Jungman:1995df,Bertone:2004pz,Feng:2010gw,Arcadi:2017kky}.

An intriguing realization of the WIMP idea is the postulation of the DM as the neutral component of an electroweak multiplet. This paradigm, commonly called minimal dark matter, was proposed long ago \cite{Cirelli:2005uq} and has been studied in detail for scalar fields~\cite{Chowdhury_2017} and fermionic fields~\cite{Hisano_2005}. 

In the case of vector DM, an effective model that included a candidate vector DM originated from a vector field in the adjoint representation of $SU(2)_L$ was first proposed in \cite{Belyaev:2018xpf}, while a complete ultraviolet model was developed in \cite{Abe_2021}. On the other hand,  models including a vector DM candidate coming from an electroweak doublet were studied in ~\cite{Saez:2018off,VanDong:2021xws}. These vector models are consistent with experimental data, if the new fields have a mass in the range of some few TeV. In this mass range, direct detection~\cite{Zurowski:2023kes} becomes difficult and discovery in colliders seems to be challenging even for future colliders \cite{Belyaev:2018xpf}. 

In these circumstances, indirect search can be a preferred alternative to discover or rule out these models.  However, it is well known that to obtain an accurate value of the DM annihilation cross section in the non-relativistic regime, it is necessary to take into account non-perturbative effects such as the so called Sommerfeld enhancement (SE) \cite{Hisano_2005}.    

The consequences of SE have been extensively studied in various scenarios, including scalar DM \cite{Chowdhury_2017, hisano2004explosive, Bellazzini:2013foa, Iengo:2009ni, Lattanzi_2009}, fermionic DM fields\cite{Hisano_2005, Biondini:2018pwp} in the doublet and triplet representations of $SU(2)_{L}$, and vector DM in the adjoint representation~\cite{Abe_2021}. In this work, we focus on computing the SE for the annihilation of vector DM in the fundamental representation of $SU(2)_{L}$ and analyzing the resulting photon flux, which is relevant for indirect detection~\cite{Zurowski:2023kes,Gaskins:2016cha, biondini2023indirect, PhysRevD.109.L041301, Becker_2022}.

A distinctive feature of our model is the presence of non-minimal interactions between the massive vector doublet and the standard gauge bosons, which introduce novel dynamics in the annihilation process. Moreover, because the DM particles in our scenario are significantly heavier than the $W$ and $Z$ bosons, the SE becomes particularly important and impactful.


Our work is organized as follows. In section \ref{sec:one}, we describe the main aspects of the model.
Section \ref{sec:trhee} is devoted to the computation of the electroweak, annihilation matrices, and annihilation cross section with SE. In Section \ref{sec:Upperl}, we compare the predictions of our model with the experimental upper limits on the annihilation cross section into $\gamma \gamma$ lines.  In section \ref{sec:Photon} we apply our results to compute the photon flux from DM annihilation at the center of the Milky Way and describe our results. Finally, in section \ref{Summary} we state our conclusions.

\section{\label{sec:one}The Model}

We start by recalling the main characteristics of the model initially developed in reference ~\citep{Saez:2018off}.  In this context,  the Standard Model (SM) is extended by the introduction of  a vector field $V_{\mu}$ which transforms as (\textbf{1}, \textbf{2}, 1/2) under the standard group $SU(3)_c \times SU(2)_L \times U(1)_Y$. This new vector field can be explicitly written as:

\begin{align}\label{vector}
V_{\mu}=\begin{pmatrix}V_{\mu}^{+}\\ V_{\mu}^{0} \end{pmatrix}=\begin{pmatrix}
 V_{\mu}^{+}\\ \dfrac{V_{\mu}^{1}+i~V_{\mu}^{2}}{\sqrt{2}}\end{pmatrix}.
\end{align}

In principle, the two  neutral  states $V_{\mu}^{1}$ and $V_{\mu}^{2}$  (the vector and axial-vector fields)  can have  different masses. However, unitarity constrains force the neutral components $V_{\mu}^{1}$ and $V_{\mu}^{2}$ to be fully degenerate ~\citep{Saez:2018off}.  In this case they combine into a complex neutral field $V_{\mu}^{0}$.
\newpage
The  Lagrangian of the model can be written as:

\begin{align}\label{Lag1}
\mathcal{L}=\mathcal{L}_{SM}&-\dfrac{1}{2}\left(D_{\mu}V_{\nu}-D_{\nu}V_{\mu} \right)^{\dagger}\left(D^{\mu}V^{\nu}-D^{\nu}V^{\mu} \right)+m_V^{2}V_{\nu}^{\dagger} V^{\nu}\nonumber\\ &-\lambda_{2}(\phi^{\dagger}\phi)(V_{\mu}^{\dagger}V^{\mu})-\lambda_{3}(\phi^{\dagger}V_{\mu})(V^{\mu\dagger}\phi)\nonumber\\ &+\lambda_{4}\left[(\phi^{\dagger}V_{\mu})(\phi^{\dagger}V^{\mu})+(V^{\mu\dagger}\phi)(V_{\mu}^{\dagger}\phi) \right]\nonumber\\ &-\alpha_{2}(V_{\mu}^{\dagger}V^{\mu})(V_{\nu}^{\dagger}V^{\nu})-\alpha_{3}(V_{\mu}^{\dagger}V^{\nu})(V_{\nu}^{\dagger}V^{\mu})\nonumber\\ &+ i\dfrac{g^{\prime}}{2}\kappa_1 V_{\mu}^{\dagger}B^{\mu\nu}V_{\nu}+ig \kappa_2 V_{\mu}^{\dagger}W^{\mu\nu}V_{\nu}, 
\end{align}

\noindent
where $B^{\mu\nu}$ represents the Abelian strength field of $U(1)_{Y}$ and $W^{\mu\nu}=W^{\mu\nu a}\frac{\tau^{a}}{2}$ is the non-Abelian strength field of $SU(2)_{L}$.  The lagrangian mass of the vector field $m_{V}$ will be corrected after the Higgs boson acquires its vacuum expectation value, due to the interaction terms between the Higgs field and the vector field present in the potential. We will call $M_V$ the physical mass of the vector boson.

Notice that the last two terms of  Lagrangian \eqref{Lag1}  describe non-minimal interactions  between the new vector field and the gauge bosons. The appearance of this  non-minimal dimension 4  terms is a special feature of  having introduced a vector doublet.  Similar non-minimal interactions in the scalar and fermion cases happen to be  higher dimensional.  In order to avoid a direct photon interaction with $V_{\mu}^{0}$,  the otherwise arbitrary constants $\kappa_1$ and $\kappa_2$  have to be equal.
For comparison purposes, we work with two values of $\kappa=\kappa_1=\kappa_2$: $\kappa=0$ and $\kappa=1$. The last value coincides with the choice made in \cite{Saez:2018off}.

Another important sector of the Lagrangian is the part of the potential which describes the interaction between the Higgs field and our massive vector doublet. Of course, the coupling constants that regulate the interactions between the Higgs field and the new vector field are free parameters. However, there is no reason to think that an important hierarchy arises among these couplings. Consequently, and for the sake of simplicity, we will take the assumption that all of them are of the same order $\lambda_2 \approx\lambda_3\approx\lambda_4\approx\lambda$. In this way, $\lambda$ will be the only constant regulating the Higgs portal. After the electroweak symmetry breaking, the physical mass of $V_{\mu}$  ($M_V$) receives contributions from the Proca term ($+m_V^{2}V_{\nu}^{\dagger} V^{\nu}$) but also from the interactions with the Higgs field. 

Finally, the $V_{\mu}$ quartic self-interactions will not contribute significantly to the SE since this phenomenon is dominated by interactions mediated by particles much lighter than $V_{\mu}$, like the electroweak gauge bosons and the Higgs boson. Consequently, we can set $\alpha_2=\alpha_3=0$.

Due to a fortunate conspiracy between the Lorentz and gauge structures  of the model, the new vector field can only appear in pairs in the Lagrangian. That means that the Lagrangian is invariant by a $Z_2$ transformation under which the new vector is odd and all the standard fields are even. Consequently, this accidental $Z_2$ symmetry makes the lighter component of $V_{\mu}$  stable. This stable component happens to be the neutral one ($V_{\mu}^{0}$).  

\section{\label{sec:trhee}Sommerfeld Enhancement Computation}
  
\subsection{Basic Formalism}

We now compute the indirect signals of $\gamma$-rays produced by DM pair annihilation in specific astrophysical environments, with a particular focus on the Galactic Center of the Milky Way. Our analysis includes the non-relativistic effect known as the Sommerfeld enhancement (SE), which accounts for the resummation of interactions between initial states. This effect can significantly amplify the annihilation cross-section, making it a crucial factor in indirect DM searches~\cite{Hisano_2005, sommerfeld1931beugung}.

Recent observational data from the Fermi Large Area Telescope (Fermi-LAT) and the Cherenkov Telescope Array (CTA) have provided stringent constraints on $\gamma$-ray signals associated with DM annihilation in the Galactic Center, offering valuable insights into DM properties and annihilation cross-sections~\cite{Silverwood_2015}. These observations underscore the importance of incorporating the SE in theoretical models to accurately predict $\gamma$-ray fluxes.

In the following paragraphs, we review the formalism used to calculate the SE ~\cite{Fujiwara:2023cns}. We then apply this framework to our model, evaluating its implications for indirect detection strategies using $\gamma$-ray observatories such as Fermi-LAT and CTA.

\begin{figure}[h]
\centering
\includegraphics[width=0.7\textwidth]{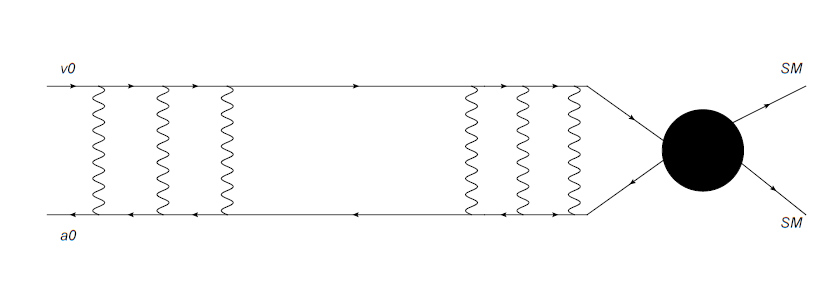}
\caption{Sommerfeld enhancement effect illustration.}
\label{Fig:Somm1}
\end{figure}
\subsubsection{The Two-Particle State and Schrödinger Equation}





 {As we have already said, the so called ``Sommerfeld Enhancement" is a phenomenon produced by long range interactions (compared to the Compton wavelength of the participants) between the initial particles in a scattering process. This is illustrated in figure \ref{Fig:Somm1}. The SE is more important in the non-relativistic regime simply because, at low velocities, the particles have more time to interact.} Since this effect is non-relativistic, we must use the Schrödinger equation for two bodies, derived from the Schwinger-Dyson equation~\cite{Hisano_2005}.  {As usual, the two body Schrödinger equation can be reduced to a one particle like equation that describes the relative motion of the participants. In natural units, The radial part of this equation can be written as:}

\begin{align}
    \left[-\dfrac{1}{M_{V}}\dfrac{d^{2}}{dr^{2}}-\frac{l(l+1)}{M_V r^2}-E+V(r)-\dfrac{i N\Gamma\delta(r)}{8\pi r^{2}}\right]g_{ab}(r)=0.
\end{align}

 The wave function $g_{ab}(r)$ describes the two particles within a doublet of $SU(2)_{L}$. The index $a$  {in the wave function $g_{ab}$} represents the initial state of the process, while $b$ corresponds to the final state. The $r$ coordinate denotes the relative position. As previously discussed, the potential $V(r)$  {represents interactions in the initial state} and originates from trilinear interactions between the new vector bosons and the electroweak bosons, including the Higgs boson. This can be expressed as $V(r)=V_{gauge}(r)+V_{Higgs}(r)$, incorporating Coulomb (long-range) and Yukawa (short-range) interactions.  {On the other hand, the annihilation matrix $\Gamma$  contains the information of the hard scattering process. In figure \ref{Fig:Somm1}, the interactions described by $\Gamma$ are represented by the dark disk. From the computational point of view, $\Gamma$ arises in the low-energy limit from effective quartic interactions between the DM and standard particles. The $N$ factor, appearing in the imaginary potential, denotes the number of spin states, given by $N=(2s+1)^2$, which in our specific case evaluates to 9. Finally, $E=\frac{M_{V}v^{2}}{4}$ is the kinetic energy of the participants and, where $v$ is their relative velocity.}

 {In principle, we should  consider all the polarized initial states in the general case}, but the main contribution in the non-relativistic limit occurs in the s-wave, i.e., when $l=0$, where the total angular momentum is given solely by the spin contribution. Therefore, the Schrödinger equation  {we need to solve} is:

\begin{align}\label{Schrödinger}
    \left[-\dfrac{1}{M_{V}}\dfrac{d^{2}}{dr^{2}}-E+V(r)-\dfrac{i N\Gamma\delta(r)}{8\pi r^{2}}\right]g_{ab}(r)=0.
\end{align}

To solve this equation \eqref{Schrödinger}, we need the boundary conditions given by \cite{Hisano_2005}:

\begin{align}
    g_{ab}(r\rightarrow 0)=\delta_{ab},
\end{align}

\begin{align}\label{Sboun}
    g_{ab}(r\rightarrow\infty)=d_{ab}(E)e^{i\sqrt{M_{V}E}r}e^{i\left(\frac{M_{V}\alpha}{2M_{V}E}\ln(2\sqrt{M_{V}E}r) \right)},
\end{align}
where $\delta_{ab}$ is a Kronecker delta, $\alpha$ is the fine-structure constant, and $d_{ab}$ are the Sommerfeld enhancement factors. The asymptotic condition \eqref{Sboun} is valid for large distances  {and includes the effects} the Coulomb and Yukawa potentials. 

 {Briefly speaking, our task consists in building the potential $V(r)$ and the annihilation matrix $\Gamma$ from the Lagrangian, to solve numerically the equation \eqref{Schrödinger}, use the asymptotic expression of $g_{ab}$ in order to identify the Somerfeld factors $d_{ab}$, and use them to calculate the average cross section $\braket{\sigma v}$.}

\subsubsection{\label{sec:four}Non-Relativistic Limit}
To ensure that the dark matter vector remains non-relativistic, we must carry out the following non-relativistic limit of the fields, as defined in~\cite{Abe_2021}.

\begin{align}\label{a17}
V_{\mu}^{0}(x)=\dfrac{1}{\sqrt{2M_V}}[v_{0}(\vec{x})\epsilon_{\mu}(x)e^{iM_V t}+a_{0}^{\dagger}(\vec{x})\epsilon_{\mu}^{\ast}(x)e^{-iM_V t}],\hspace{0.5cm}
\end{align}

\begin{align}\label{a18}
(V_{\mu}^{0})^{\ast}(x)=\dfrac{1}{\sqrt{2M_V}}[a_{0}(\vec{x})\epsilon_{\mu}(x)e^{iM_V t}+v_{0}^{\dagger}(\vec{x})\epsilon_{\mu}^{\ast}(x)e^{-iM_V t}],\hspace{0.5cm}
\end{align}

\begin{align}\label{a19}
V_{\mu}^{+}(x)=\dfrac{1}{\sqrt{2M_V}}[v_{+}(\vec{x})\epsilon_{\mu}(x)e^{iM_V t}+v_{-}^{\dagger}(\vec{x})\epsilon_{\mu}^{\ast}(x)e^{-iM_V t}],\hspace{0.5cm}
\end{align}

\begin{align}\label{a20}
V_{\mu}^{-}(x)=\dfrac{1}{\sqrt{2M_V}}[v_{+}^{\dagger}(\vec{x})\epsilon_{\mu}^{\ast}(x)e^{-iM_V t}+v_{-}(\vec{x})\epsilon_{\mu}(x)e^{iM_V t}],\hspace{0.5cm}
\end{align}

Here, $v_{0},a_{0},v_{+},v_{-}$ represent the non-relativistic limits of the DM vector field. $M_V$ denotes the mass of the neutral component of the DM vector field, and $\epsilon_{\mu}$ signifies the polarization vector. These polarization vectors adhere to the transverse conditions and an orthonormal relation:

\begin{align}
  p^{\mu}\epsilon_{\mu}=0; \hspace{1.0cm} \epsilon_{\mu}\epsilon_{\nu}=-\delta_{\mu\nu}.
\end{align}

\subsection{\label{sec:five}Electroweak Matrix}
\subsubsection{Gauge Bosons-Dark Matter Interaction}

We aim to investigate the annihilation of DM particles into SM ones. Specifically, we consider the pairs ($v_{0}v_{0}$), for which such annihilation occurs. The most important contributions come from processes involving the exchange of $\gamma$, $Z$, $W^{\pm}$ bosons, having in their initial states ($v_{0}v_{0}$) , ($v_{0}a_{0}$), or ($v_{+}v_{-}$)  pairs. Our approach involves deriving an effective action that accurately describes these pairs. For this purpose, we define the following vector:
\begin{align}\label{vectornonrel}
s(x,\vec{y})=\begin{pmatrix}\dfrac{v_{0}(x)v_{0}(x^{0},\vec{y})}{\sqrt{2}}\\ \dfrac{a_{0}(x)a_{0}(x^{0},\vec{y})}{\sqrt{2}} \\v_{0}(x)a_{0}(x^{0},\vec{y}) \\ v_{+}(x)v_{-}(x^{0},\vec{y})
\end{pmatrix}.
\end{align}

The disparity in the normalization of the first component in \eqref{vectornonrel} arises from the fact that they pertain to states composed of two identical particles. The effective action is as follows:

\begin{align}\label{effectiveact}
S_{eff,1}=-\int&\dfrac{d^{4}xd^{3}y}{8\pi |\vec{x}-\vec{y}|}   (J_{A}^{0}(x)J_{A}^{0}(x^{0},\vec{y})\nonumber\\
&+J_{Z}^{0}(x)J_{Z}^{0}(x^{0},\vec{y})e^{-M_{Z}|\vec{x}-\vec{y}|}\nonumber\\
&+2J_{W_{+}}^{0}(x)J_{W_{-}}^{0}(x^{0},\vec{y})e^{-M_{W}|\vec{x}-\vec{y}|}).  
\end{align}  

The currents are calculated from the cubic interactions of the Lagrangian~\eqref{Lag1} between DM and gauge bosons. From these expressions, we employ the non-relativistic limit to obtain the electroweak matrix, which we will use to solve the Schrödinger equation~\eqref{Schrödinger}.

In the non-relativistic limit, the first term of equation \eqref{effectiveact} characterizes the Coulomb potential between non-relativistic particles, whereas the remaining terms depict the Yukawa potentials  related to the exchange of $Z$ and $W$ bosons. Upon employing vector \eqref{vectornonrel}, the effective action transforms into the following expression:

\begin{align}\label{effactnonrel}
S_{eff,1_{NR}}=-\int d^{4}xd^{3}y~s(x,\vec{y})^{\dagger}V_{g-V}(|\vec{x}-\vec{y}|)s(x,\vec{y}),
\end{align}

where $V_{g-V}(|\vec{x}-\vec{y}|)$ is electroweak potential matrix (interaction gauge boson-vector DM), with $r=|\vec{x}-\vec{y}|$;

\begin{align}\label{102}
V_{g-V}(r)=-\dfrac{g^{2}}{8\pi r} \begin{pmatrix}4a^{2}e^{-M_{Z}r} & 0 & 0 &0\\
0 & 4a^{2}e^{-M_{Z}r} & 0 & 0\\
0 & 0 & 16a^{2}e^{-M_{Z}r} & b^{2}e^{-M_{W}r}\\
0& 0 & b^{2}e^{-M_{W}r} &8a^{2}e^{-M_{Z}r}-2(c_{W}(1-(\kappa_{1}+\kappa{2})-s_{W}t_{W}^{2})^{2} 
\end{pmatrix},
\end{align}
with $a=(c_{W}(\kappa_{1}+\kappa_{2})+s_{W})$, $b=(\frac{3}{8}-2\sqrt{2})(\kappa_{1}+\kappa_{2})+2\sqrt{2}$
In appendix \ref{A}, we show how to compute this matrix.

\subsubsection{Higgs Boson-Dark Matter Interaction}
Now we need to take into account the interaction between the Higgs boson and the DM vector. Our primary focus lies in understanding the cubic interaction.
For this purpose, we parameterize the Higgs doublet as follows:

\begin{align}\label{gold}
\phi(x) =\begin{pmatrix} \phi^{+}(x)\\ \dfrac{v+h(x)}{\sqrt{2}}+i\phi^{0}(x)
\end{pmatrix},
\end{align}
where  $\phi^{+}$ and  $\phi^{0}$ are would-be Goldtone bosons, and  $v$ is vacuum expectation value.

Now, when we substitute $V_{\mu}$, $V_{\mu}^{\dagger}$, $\phi$, and $\phi^{\dagger}$ into the Lagrangian \eqref{Lag1}, we can derive the electroweak matrix describing the interaction between the Higgs and the DM vector, which takes the form:

\begin{align}\label{V h-V}
V_{h-V}(r)=\dfrac{v^{2}}{32\pi M_V^{2} r}\begin{pmatrix}
8\lambda_{4}^{2}a & 0 & 0 & 0\\
0 & 8\lambda_{4}^{2}a & 0 & 0\\
0 & 0 & \dfrac{1}{2} (\lambda_{2}+\lambda_{3})^{2}a & (\lambda_{3}+2\lambda_{4})^{2}e^{-M_{W}r}\\
0 & 0 & (\lambda_{3}+2\lambda_{4})^{2}e^{-M_{W}r} & \lambda_{2}^{2}a
\end{pmatrix},
\end{align}

with $a=(2e^{-M_{h}r}+e^{-M_{Z}r})$. In Appendix~\ref{A}, we illustrate the methodology for computing this matrix.

\subsection{\label{sec:six}Annihilation Matrix}

\subsubsection{Gauge Boson-Dark Matter Vector}

When DM particles annihilate into SM ones, it becomes essential to compute the annihilation matrix, specifically examining the interaction between gauge bosons and the DM. This requires the use of the quartic terms from Lagrangian \eqref{Lag1}. The effective action for these quartic terms, in the non-relativistic limit, can be expressed as follows:

\begin{align}\label{acc2}
S_{eff,2 NR}&=2i\int ~d^{4}x d^{3}y~ s(x,\Vec{y})^{\dagger}\Gamma_{gauge}\delta^{3}(|\Vec{x}-\Vec{y}|) s(x,\Vec{y}).
\end{align}

In Appendix \ref{B}, the computation of these terms is detailed. Considering the different channels and total spin states, we obtain the following annihilation matrices:

\begin{align}\label{mww}
\Gamma_{WW}^{J=0}=\dfrac{2g^{4}}{64\sqrt{3}\pi M_V^{2}}\begin{pmatrix}
2 & 0 & 0 & 0\\
0 & 2 & 0 & 0\\
0 & 0 & 1 & 1\\
0 & 0 & 1 & 1
\end{pmatrix},
\end{align}

\begin{align}\label{MatrizW1}
    \Gamma_{WW}^{J=2}=\dfrac{g^{4}}{64\pi M_V^{2}}\left(\dfrac{1+3\cos(2\theta)}{2\sqrt{6}}+\sin^{2}\theta \right)\begin{pmatrix}
2 & 0 & 0 & 0\\
0 & 2 & 0 & 0\\
0 & 0 & 1 & 1\\
0 & 0 & 1 & 1
\end{pmatrix},
\end{align}

\begin{align}\label{maa1}
\Gamma_{AA}^{J=0}=\dfrac{e^{4}}{128\sqrt{3}\pi M_V^{2}}\begin{pmatrix}
2a^{2} & 0 & 0 & 0\\
0 & 2a^{2} & 0 & 0\\
0 & 0 & a^{2} & 4a^{2}\\
0 & 0 & 4a^{2} & 4
\end{pmatrix},
\end{align}

\begin{align}\label{mma2}
    \Gamma_{AA}^{J=2}=\dfrac{e^{4}}{256\pi M_V^{2}}\left(\dfrac{1+3\cos(2\theta)}{2\sqrt{6}}+\sin^{2}\theta \right)\begin{pmatrix}
2a^{2} & 0 & 0 & 0\\
0 & 2a^{2} & 0 & 0\\
0 & 0 & a^{2} & 4a^{2}\\
0 & 0 & 4a^{2} & 4
\end{pmatrix},
\end{align}
with $a=(1-\tan\theta_{W})$.

\begin{align}\label{mzz}
\Gamma_{ZZ}^{J=0}&=\dfrac{g^{4}}{128\sqrt{3}\pi M_V^{2}}\begin{pmatrix}
2b^{2} & 0 & 0 & 0\\
0 & 2b^{2} & 0 & 0\\
0 & 0 & 2b^{2} & -4c\\
0 & 0 & -4c & 4\sin^{4}\theta_{W}
\end{pmatrix},\\\Gamma_{ZZ}^{J=2}&=\dfrac{g^{4}}{256\pi M_V^{2}}\left(\dfrac{1+3\cos(2\theta)}{2\sqrt{6}}+\sin^{2}\theta \right)\begin{pmatrix}
2b^{2} & 0 & 0 & 0\\
0 & 2b^{2} & 0 & 0\\
0 & 0 & 2b^{2} & -4c\\
0 & 0 & -4c & 4\sin^{4}\theta_{W}
\end{pmatrix},\nonumber
\end{align}

with $b=(\tan^{2}\theta_{W}\sin^{2}\theta_{W}-\sin\theta_{W}\cos\theta_{W})$ and $c=\sin^{2}\theta_{W}\tan\theta_{W}(\sin^{2}\theta_{W}-\cos^{2}\theta_{W})$.

\begin{align}\label{maz}
\Gamma_{AZ}^{J=0}&=\dfrac{e^{2}g^{2}}{128\sqrt{3}\pi M_V^{2}}\begin{pmatrix}
d^{2} & 0 & 0 & 0\\
0 & d^{2} & 0 & 0\\
0 & 0 & d^{2} & f\\
0 & 0 & f & \dfrac{\sin^{2}(2\theta_{W})}{\cos^{2}\theta_{W}}
\end{pmatrix},\\\Gamma_{AZ}^{J=2}&=\dfrac{e^{2}g^{2}}{256\pi M_V^{2}}\left(\dfrac{1+3\cos(2\theta)}{2\sqrt{6}}+\sin^{2}\theta \right)\begin{pmatrix}
d^{2} & 0 & 0 & 0\\
0 & d^{2} & 0 & 0\\
0 & 0 & d^{2} & f\\
0 & 0 & f & \dfrac{\sin^{2}(2\theta_{W})}{\cos^{2}\theta_{W}}
\end{pmatrix},\nonumber
\end{align}

with $d=(\cos\theta_{W}-\sin\theta_{W})$ and $f=(\cos\theta_{W}-\sin\theta_{W})^{4}(1+\tan\theta_{W})^{2}$.

\subsubsection{Higgs Boson-Dark Matter Vector}

We follow a similar procedure to take into accounts the Higgs boson and the DM  interactions. In this case, we require the quartic terms from Lagrangian \eqref{Lag1}, the quartic terms involving the new vector field and the Higgs boson doublet. The resulting annihilation matrix can be written as follows:

\begin{align}\label{ani h-V}
\Gamma_{Higgs}=\dfrac{1}{32\pi M_V^{2}}\begin{pmatrix}
\left(\dfrac{3}{2}+\sqrt{2} \right)^{2}\lambda_{4}^{2} & 0 & 0 & 0\\
0 & \dfrac{9}{4}\lambda_{4}^{2} & 0 & 0\\
0 & 0 & \dfrac{1}{4}(7\lambda_{2}^{2}+6\lambda_{2}\lambda_{3}+5\lambda_{3}^{2}) & \dfrac{1}{2}(\lambda_{3}+2\lambda_{4})^{2}\\
0 & 0 & \dfrac{1}{2}(\lambda_{3}+2\lambda_{4})^{2} & \dfrac{3}{4}\lambda_{2}^{2}+(\lambda_{2}+\lambda_{3})^{2}
\end{pmatrix}.
\end{align}

\subsection{Numerical Results}

Now, we have all the needed ingredients to compute the thermal averaged cross-section $\braket{\sigma v}$.  {This is done by using the optical theorem, which relates the imaginary part of the Green's functions to the thermally averaged total annihilation cross-section~\cite{Hisano_2005,Abe_2021} of DM particles:

\begin{align}\label{cross2}
 \braket{\sigma v}_{XX^{\prime}}= c\sum_{a,b}\sum_{J,J_{z}}(\Gamma_{XX^{\prime}}^{J,J_{z}})_{ab}~d_{3 a}(E)d_{3 b}^{\ast}(E),
\end{align}

where $c$ is a normalization constant (2 for $a=b$, 1 for $a\neq b$). The fixed index 3 is related to the definition of the vector $\vec{s}$ \eqref{vectornonrel}, to select dark matter annihilation $(V_{0}V_{0} \longrightarrow XX^{\prime})$ with $X$ and $X'$ being particles of the SM.}

\begin{figure}
 \centering
  \subfloat[$\lambda=0$]{
   \label{SE0kappa}
   \includegraphics[width=0.55\textwidth]{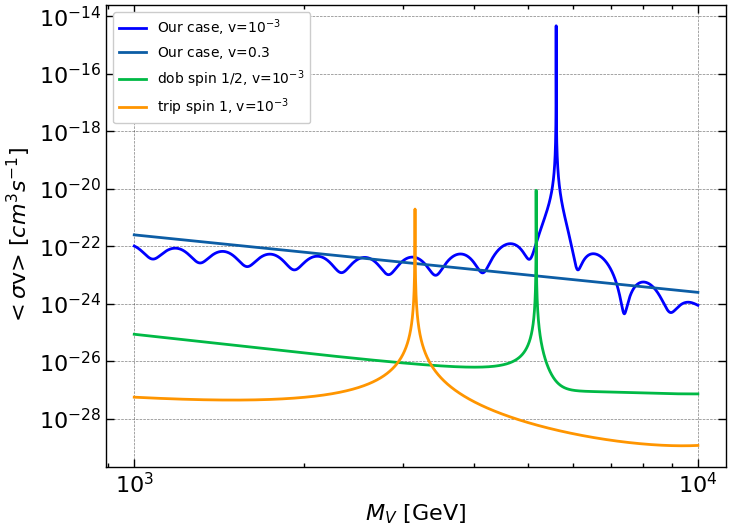}}
  \subfloat[$\lambda=0.5$]{
   \label{SE005kappa}
    \includegraphics[width=0.55\textwidth]{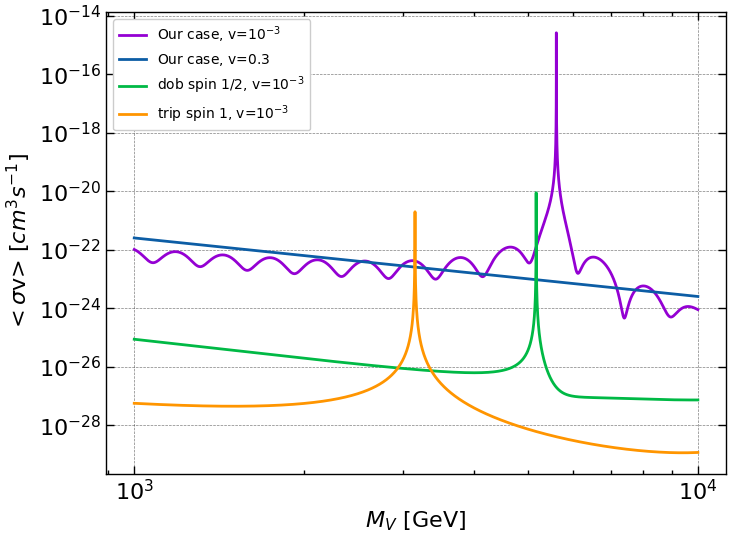}}\\
   \subfloat[$\lambda=2$]{
   \label{SE2kappa}
    \includegraphics[width=0.55\textwidth]{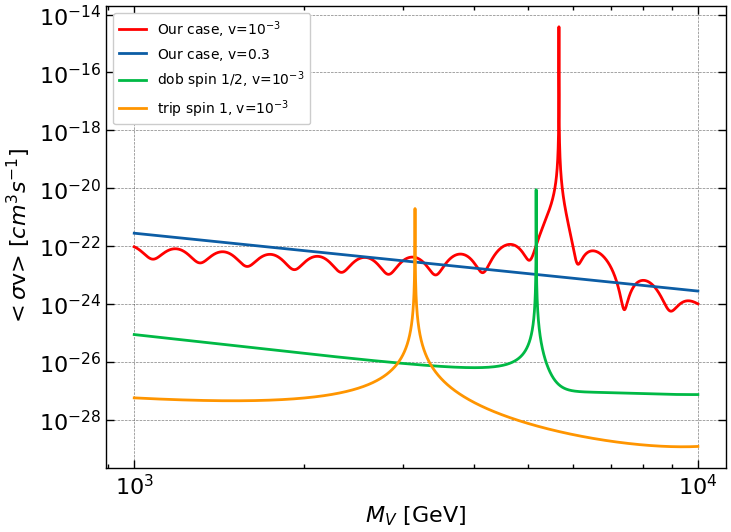}}
    \subfloat[$\lambda=9$]{
   \label{SE7kappa}
    \includegraphics[width=0.55\textwidth]{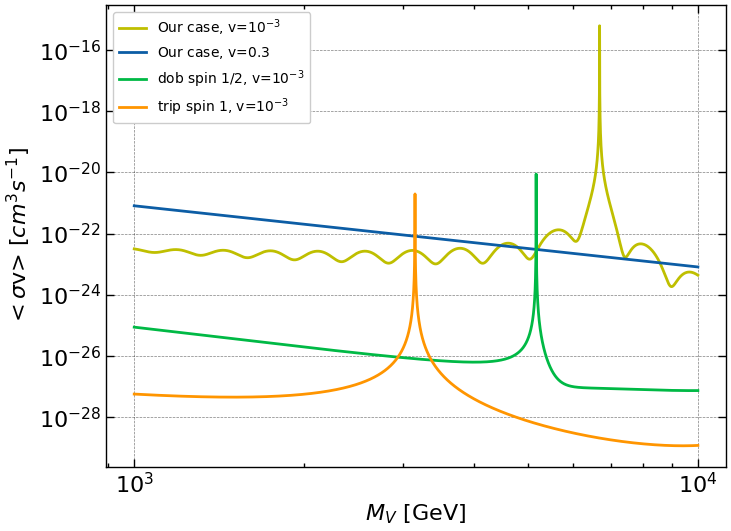}}
 \caption{Annihilation cross-section (fundamental representation spin-1 of $SU(2)_{L}$) for the process $v_{0}a_{0}\longrightarrow SMSM$ with the Sommerfeld Enhancement factor, considering velocities $v=10^{-3}$ (galactic Halo, shown as continuous lines, blue, purple, red and yellow lines) and $v=0.3$ (Relic density, indicated by blue continuous line without resonance), and different values of the Higgs portal coupling constant $\lambda$, with $\kappa_{1}=\kappa_{2}=0$. The orange line indicates that it is the spin-1 triplet representation of $SU(2)_{L}$, while the green line indicates that it is the spin-1/2 doublet representation of $SU(2)_{L}$.}
 \label{SommerKappa}
\end{figure}

\begin{figure}
 \centering
  \subfloat[$\lambda=0$]{
   \label{SE0}
   \includegraphics[width=0.55\textwidth]{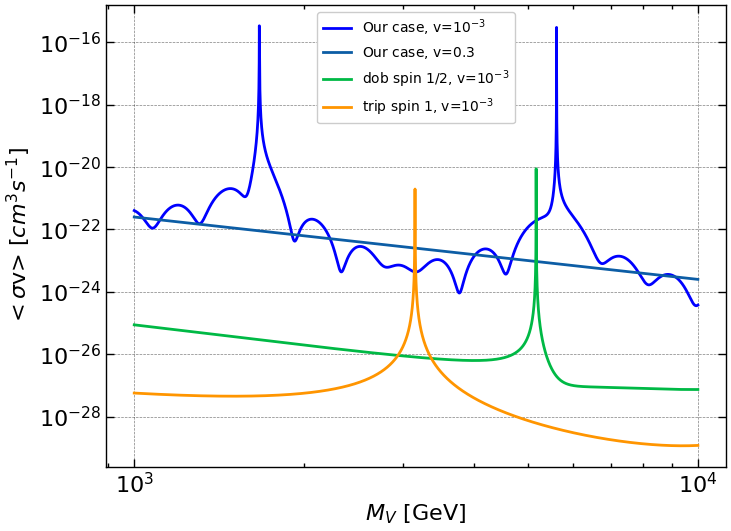}}
  \subfloat[$\lambda=0.5$]{
   \label{SE005}
    \includegraphics[width=0.55\textwidth]{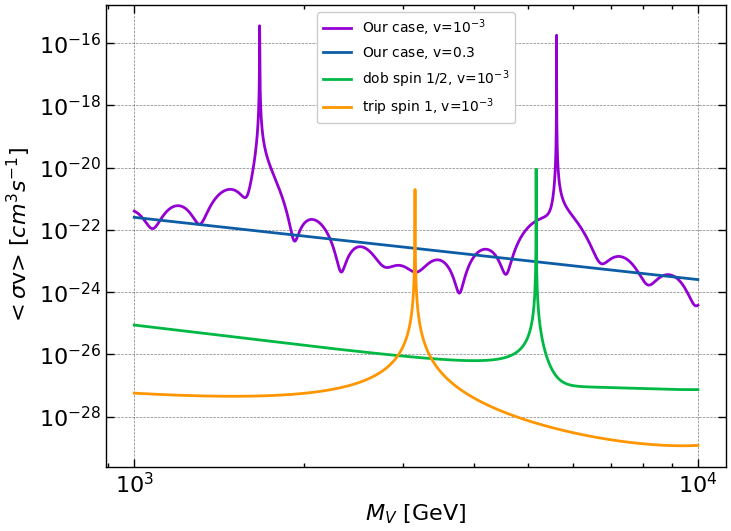}}\\
   \subfloat[$\lambda=2$]{
   \label{SE2}
    \includegraphics[width=0.55\textwidth]{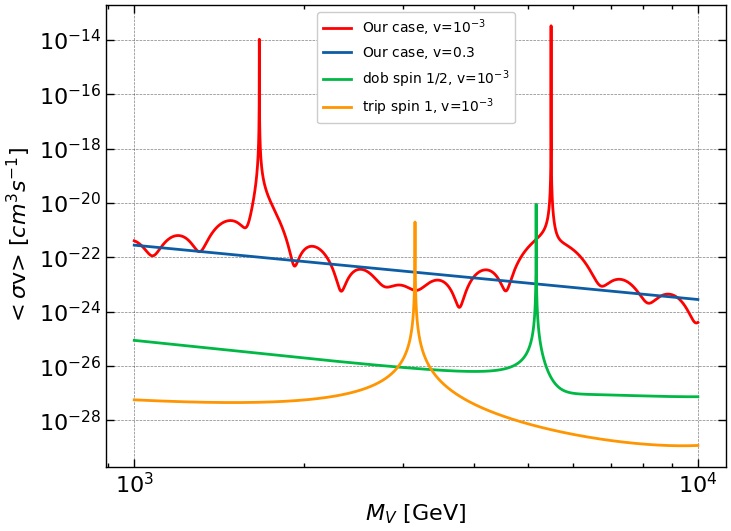}}
    \subfloat[$\lambda=9$]{
   \label{SE2kappa1}
    \includegraphics[width=0.55\textwidth]{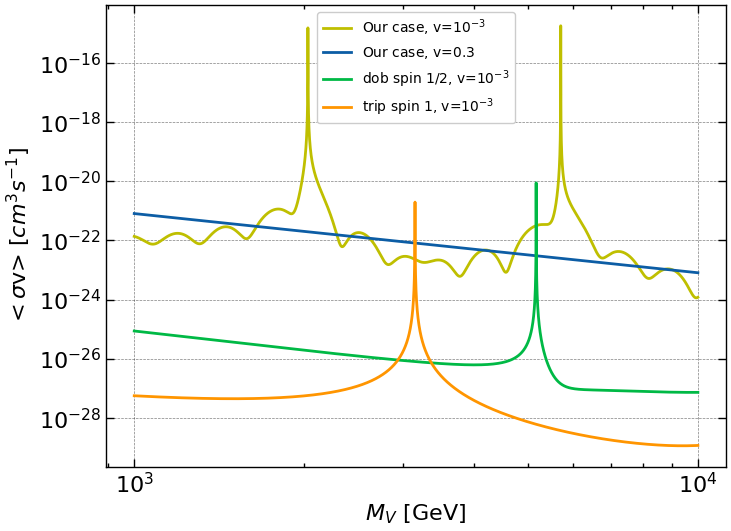}}
 \caption{Annihilation cross-section (fundamental representation spin-1 of $SU(2)_{L}$) for the process $v_{0}a_{0}\longrightarrow SMSM$ with the Sommerfeld Enhancement factor, considering velocities $v=10^{-3}$ (galactic Halo, shown as continuous lines, blue, purple, red and yellow lines) and $v=0.3$ (Relic density, indicated by blue continuous line without resonance), and different values of the Higgs portal coupling constant $\lambda$, with $\kappa_{1}=\kappa_{2}=1$. The orange line indicates that it is the spin-1 triplet representation of $SU(2)_{L}$, while the green line indicates that it is the spin-1/2 doublet representation of $SU(2)_{L}$.}
 \label{Sommer}
\end{figure}

When we expand over all the annihilation channels,  $\braket{\sigma v}$ takes the form:

\begin{align}\label{SETOT}
\braket{\sigma v}=\braket{\sigma v}_{W^{+}W^{-}}+\braket{\sigma v}_{ZZ}+\braket{\sigma v}_{\gamma\gamma}+\braket{\sigma v}_{Z\gamma}+\braket{\sigma v}_{hh}.   
\end{align}
Our results are presented in Figures \ref{SommerKappa} and \ref{Sommer}. The first figure displays our predictions without considering the non-minimal interactions $(\kappa=0)$, while the second figure incorporates these interactions with $(\kappa=1)$. In both cases, several notable features of the Sommerfeld enhancement (SE) emerge that warrant discussion.

Firstly, as is well-known, SE is a velocity-dependent phenomenon. For instance, at a relative velocity of $v=0.3$ (dark blue continuous lines in Figures \ref{SommerKappa} and \ref{Sommer}), which corresponds to the conditions of the early Universe when the relic density was established, no significant structure is observed. 

In contrast, for smaller velocities such as those expected within the galaxy, $v=10^{-3}$, pronounced oscillations and resonances become evident. We observe that the small oscillations depend on interactions mediated by the Higgs boson. Notably, these oscillations are attenuated for large values of the portal coupling constant $\lambda$. Interestingly, the inclusion of non-minimal interactions ($\kappa\neq 0$) introduces an additional resonance at a lower mass. However, the $\kappa$-terms do not significantly alter the magnitude of the oscillations or the amplitude of the resonances.

For masses that exhibit such resonances, approximately around $1.91$ TeV and $6$ TeV, the SE factor experiences a substantial boost, increasing by nearly five orders of magnitude across all values of the portal coupling constant $\lambda$. This remarkable enhancement underscores the critical role of these parameters in shaping the indirect detection prospects for dark matter.


\newpage
\section{\label{sec:Upperl}Gamma-ray line constraints}
\begin{figure}
 \centering
 \includegraphics[width=0.67\textwidth]{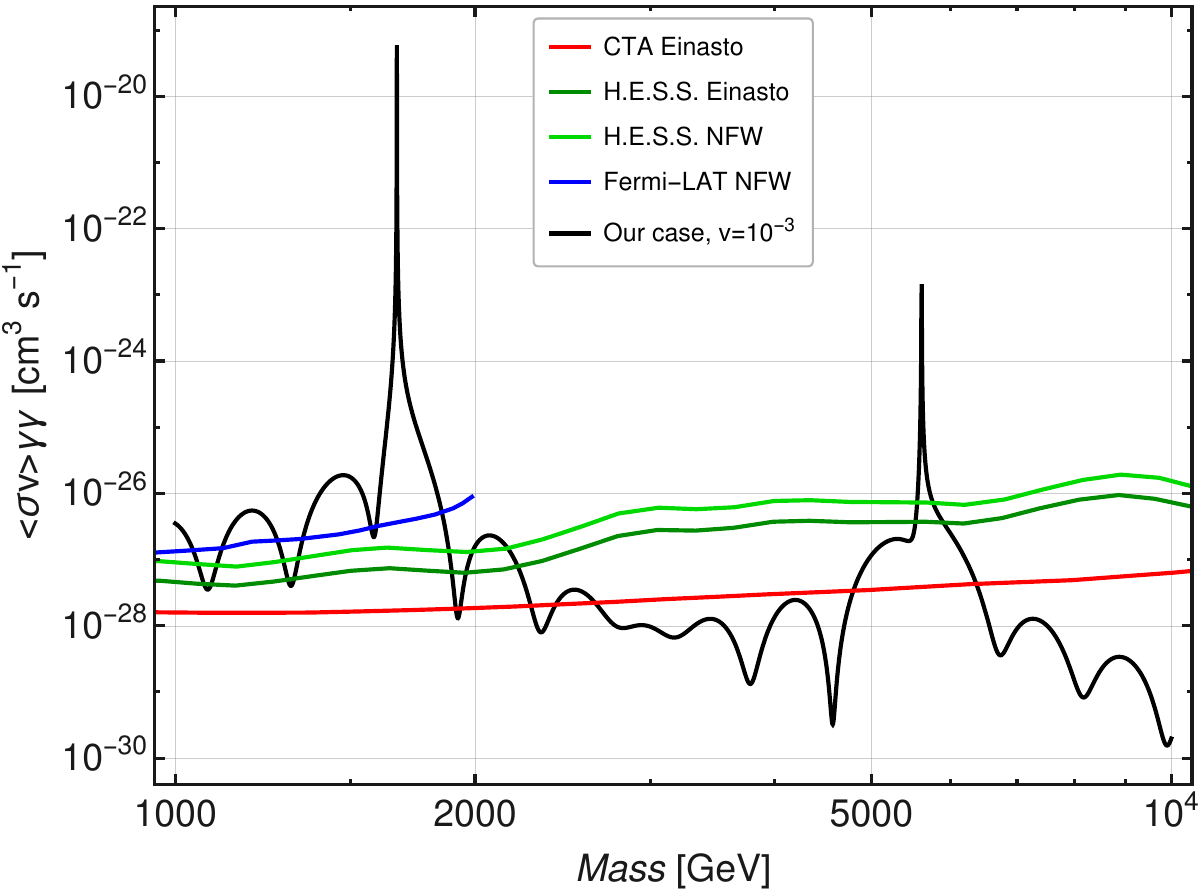}
 \caption{Comparison between the experimental upper limits on the annihilation cross section $\langle \sigma v \rangle_{\gamma \gamma}$ and the predictions of our model ($SU(2)_{L}$ vector doublet). The solid blue, green, and red lines show the limits from the Fermi-LAT, H.E.S.S., and the projected sensitivities of the CTA, respectively, assuming Einasto and NFW halo profiles. The solid black line corresponds to the prediction of our model for $\lambda=5$ and $\kappa_{1}=\kappa_{2}=1$, considering the process $v_{0}a_{0}\rightarrow \gamma \gamma$ with the Sommerfeld enhancement factor and $v=10^{-3}$.}
 \label{figUP}
\end{figure}

A monochromatic gamma-ray line signal constitutes one of the most distinctive signatures of the presence of DM \cite{Bringmann:2012ez,Fermi-LAT:2013thd,HESS:2018cbt}. This type of signal can arise from DM annihilation processes at loop level, in which a photon is produced together with another neutral particle ($Y=\gamma,Z,h$). In particular, if $Y=\gamma$, the photon energy coincides with the DM particle mass, $E=m_{DM}$.

In our case, for Figure \ref{figUP} we consider the model predictions for the channel $\gamma \gamma$, comparing them with the corresponding experimental limits on the annihilation cross section $\langle \sigma v \rangle_{\gamma \gamma}$. These limits are derived from observations carried out by the Fermi-LAT and H.E.S.S. telescopes, together with the projected sensitivities of CTA. In particular, we use 14 years of Fermi-LAT $\gamma$-ray data in the energy range from $10$ GeV to $2$ TeV \cite{Foster:2022nva}, 10 years of H.E.S.S. data in the energy window between $300$ GeV and $70$ TeV \cite{HESS:2018cbt}, and the projected CTA sensitivity for an observation time of 500 hours in the range from $200$ GeV to $100$ TeV \cite{CTAO:2024wvb}, considering regions of interest around the Galactic Center. Since the energy range relevant for the process $v_{0}a_{0}\rightarrow \gamma \gamma$, lies between $1$ and $10$ TeV, we restrict Figure \ref{figUP} to this interval.

The most stringent constraints currently come from the H.E.S.S. telescope, with the Einasto density profile imposing stronger limits than the NFW profile, and being surpassed only by the projected sensitivities of the CTA. Two clearly excluded regions are identified where the model exceeds the experimental limits, which coincide with the resonances associated with the SE, around $1.8$ TeV and $6$ TeV. 

On the other hand, in the region $\langle \sigma v \rangle_{\gamma \gamma} \sim 10^{-30} - 10^{-27}\ \text{cm}^3\text{s}^{-1}$, our model lies below the experimental limits over a large portion of the mass range, defining a wide allowed region. However, for considerably smaller values of the annihilation cross section, the resulting $\gamma$-ray signal would be too weak to be detected by current instruments. Nevertheless, there exists a band in the parameter space, still compatible with the limits from the Fermi-LAT and the H.E.S.S., and lying above the projected sensitivity of CTA ($\langle \sigma v \rangle_{\gamma \gamma} \sim 10^{-28}\ \text{cm}^3\text{s}^{-1}$), which is particularly relevant from an experimental perspective. In this region, the model predicts signals potentially accessible to future observations, positioning CTA as a key instrument to probe this scenario in the TeV scale mass range.

\newpage
\section{\label{sec:Photon}Photons Flux from Annihilation neutral Dark Matter particles}

The $\gamma$-ray flux resulting from the annihilation of neutral DM particles, as described in~\cite{PhysRevD.109.L041301}, in a region of the sky with a solid angle $\Omega$ is given by:

\begin{align}\label{diffphotons}
\dfrac{d\Phi}{dE_{\gamma}}=\sum_{i}\dfrac{1}{8\pi}J(\Omega)\dfrac{\braket{\sigma v}_{i}}{M_V^{2}}\left(\dfrac{dN_{i}}{dE_{\gamma}} \right).    
\end{align}
 Here, $J(\Omega)$  represents the astrophysical $J$-factor, which encodes the DM distribution in the observed object, and is defined as

\begin{align}\label{Jastro}
J(\Omega)=\dfrac{1}{\Delta\Omega}\int d\Omega\int d\ell~ \rho[r(\ell,\psi)]^{2},   
\end{align}

where $\rho[r(\ell,\psi)]$ is the DM density, which depends on the assumed astrophysical profile. The values of the $J$-factor span several orders of magnitude depending on the astrophysical target and the adopted region of observation. For DM annihilation, integrating over a circular region with an angular radius of $0.5^{\circ}$ yields characteristic values of $J_{\rm ann}\sim 10^{19}\, \rm{GeV}^{2}/cm^{5}$ for dwarf spheroidal galaxies, $J_{\rm ann}\sim 10^{20}\, \rm{GeV}^{2}/cm^{5}$ for the Andromeda galaxy, and  $J_{\rm ann}\sim 10^{21}\, \rm{GeV}^{2}/cm^{5}$ for the Galactic Center of the Milky Way \cite{Gaskins:2016cha,Fornasa:2013iaa,Geringer-Sameth:2014yza,Sanchez-Conde:2011zys}.


The distribution, denoted as $\left(\frac{dN_{i}}{dE_{\gamma}}\right)$, corresponds to the photon spectrum and describes the number of photons produced per unit energy $E_{\gamma}$, while $\langle \sigma v \rangle_{i}$ represents the thermally averaged annihilation cross section of the i-th channel.

Taking into account the results of Section \ref{sec:Upperl}, we compute the differential photon flux predicted by our model using equation \eqref{diffphotons}. To this end, we use the values of $\langle \sigma v \rangle$ previously obtained from equation \eqref{cross2}, together with the spectral tables from \cite{Marco_Cirelli_2011} and tools available in \cite{PPPC4DMID_web}. In addition, we consider the Galactic Center as the region of interest, with a solid angle of $\Delta\Omega=0.96\times10^{-3}$ sr (see Table 2 in \cite{Cirelli:2010xx}), corresponding to an angular aperture of $1^{\circ}$.

\begin{figure}
 \centering
  \subfloat[$v_{0}a_{0}\longrightarrow\gamma\gamma$]{
   \label{flux0kappa}
    \includegraphics[width=0.5\textwidth]{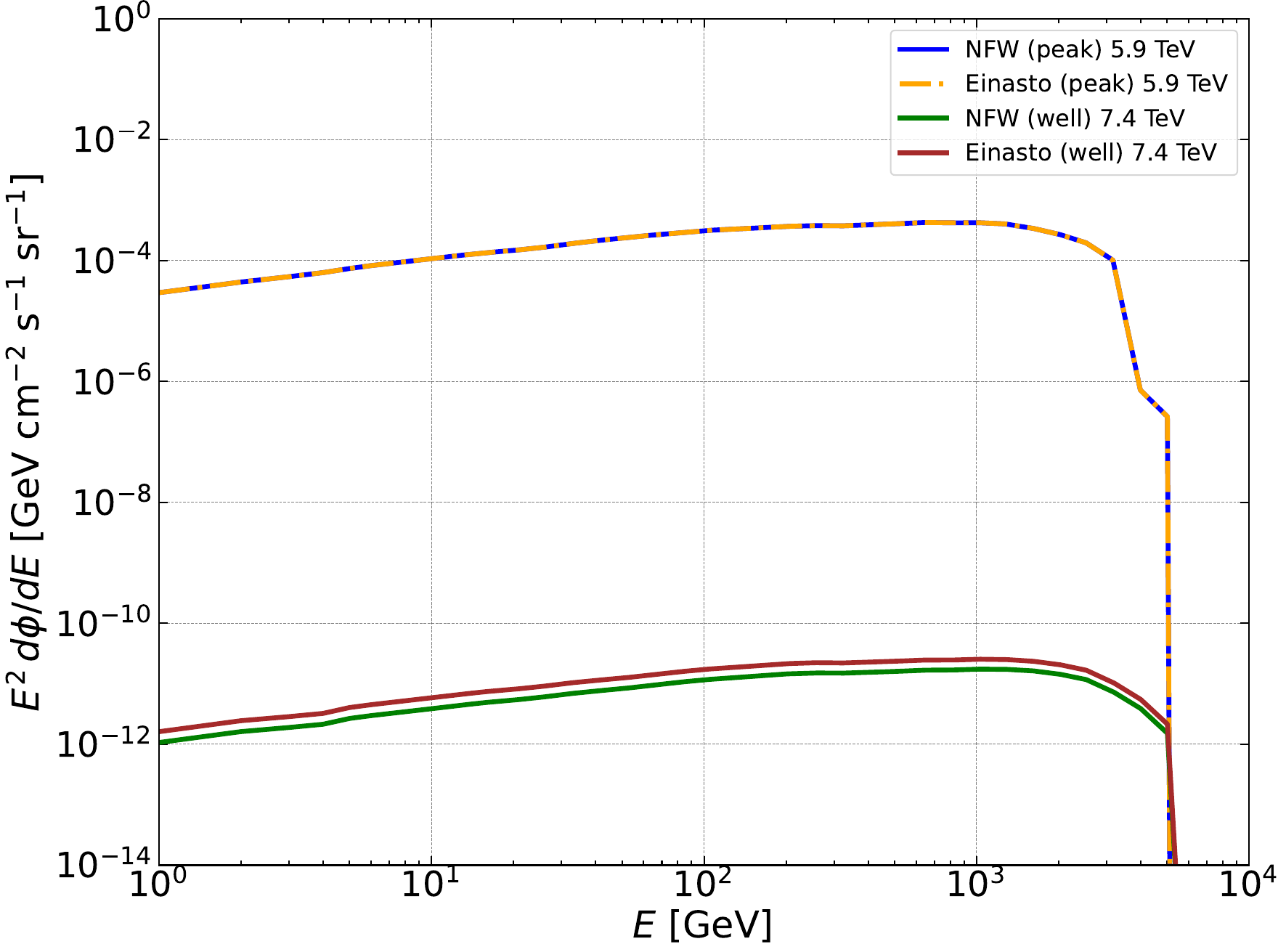}}
  \subfloat[$v_{0}a_{0}\longrightarrow W^{+}W^{-}$]{
   \label{flux0.5kappa}
    \includegraphics[width=0.5\textwidth]{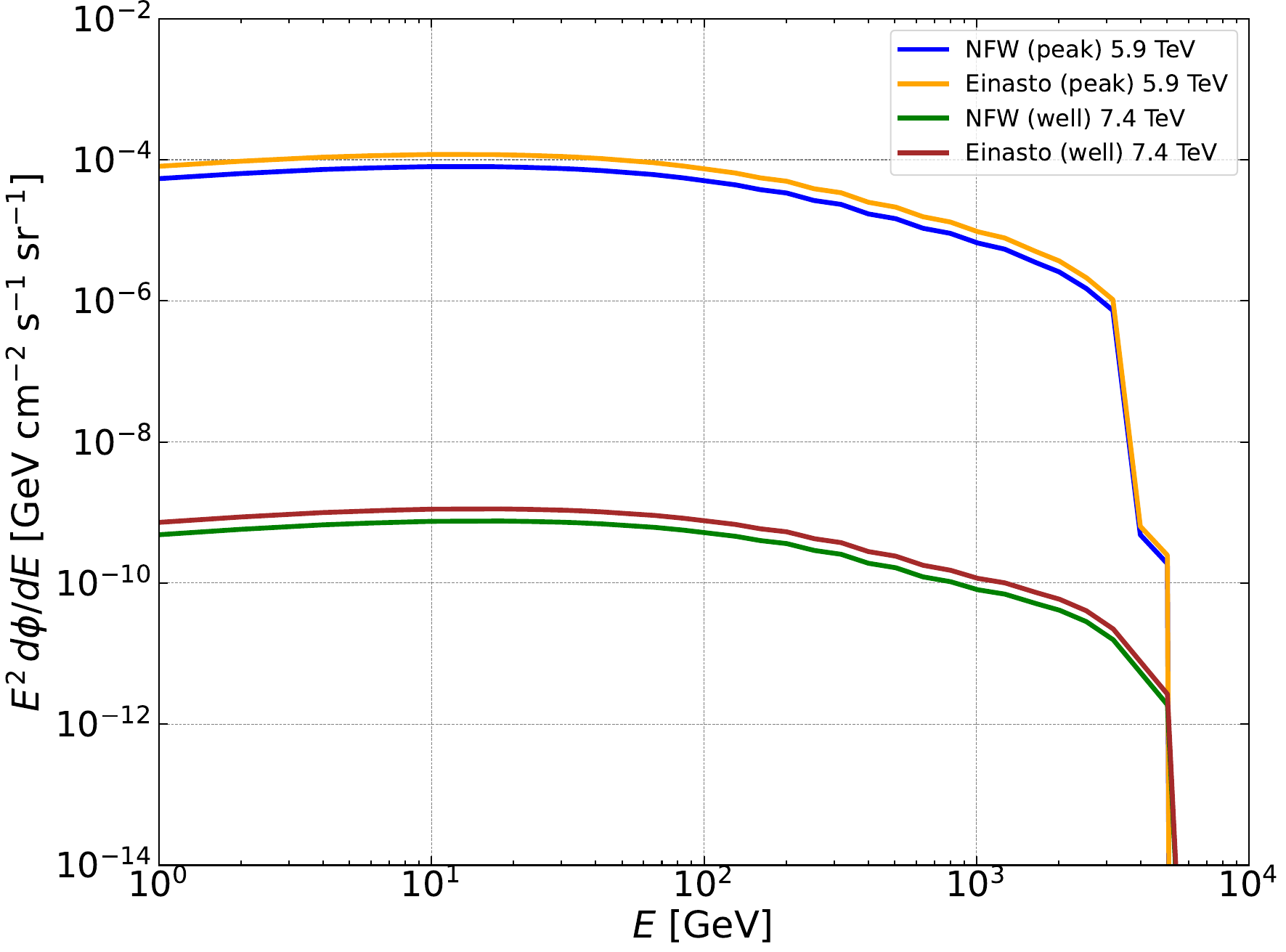}}\\
    \subfloat[$v_{0}a_{0}\longrightarrow ZZ$]{
   \label{flux2kappa}
    \includegraphics[width=0.5\textwidth]{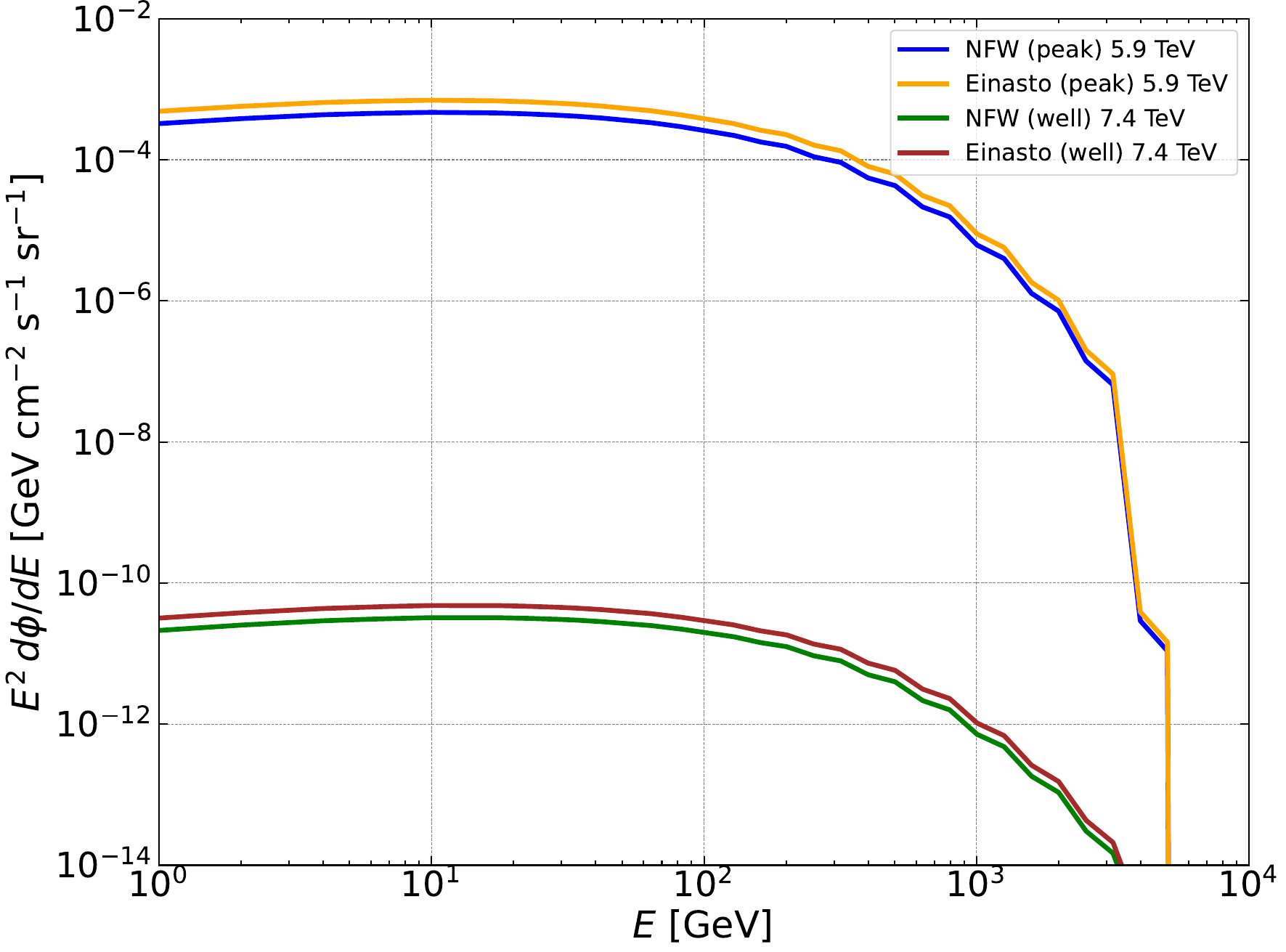}}
  \subfloat[$v_{0}a_{0}\longrightarrow hh$]{
   \label{flux4pikappa}
    \includegraphics[width=0.5\textwidth]{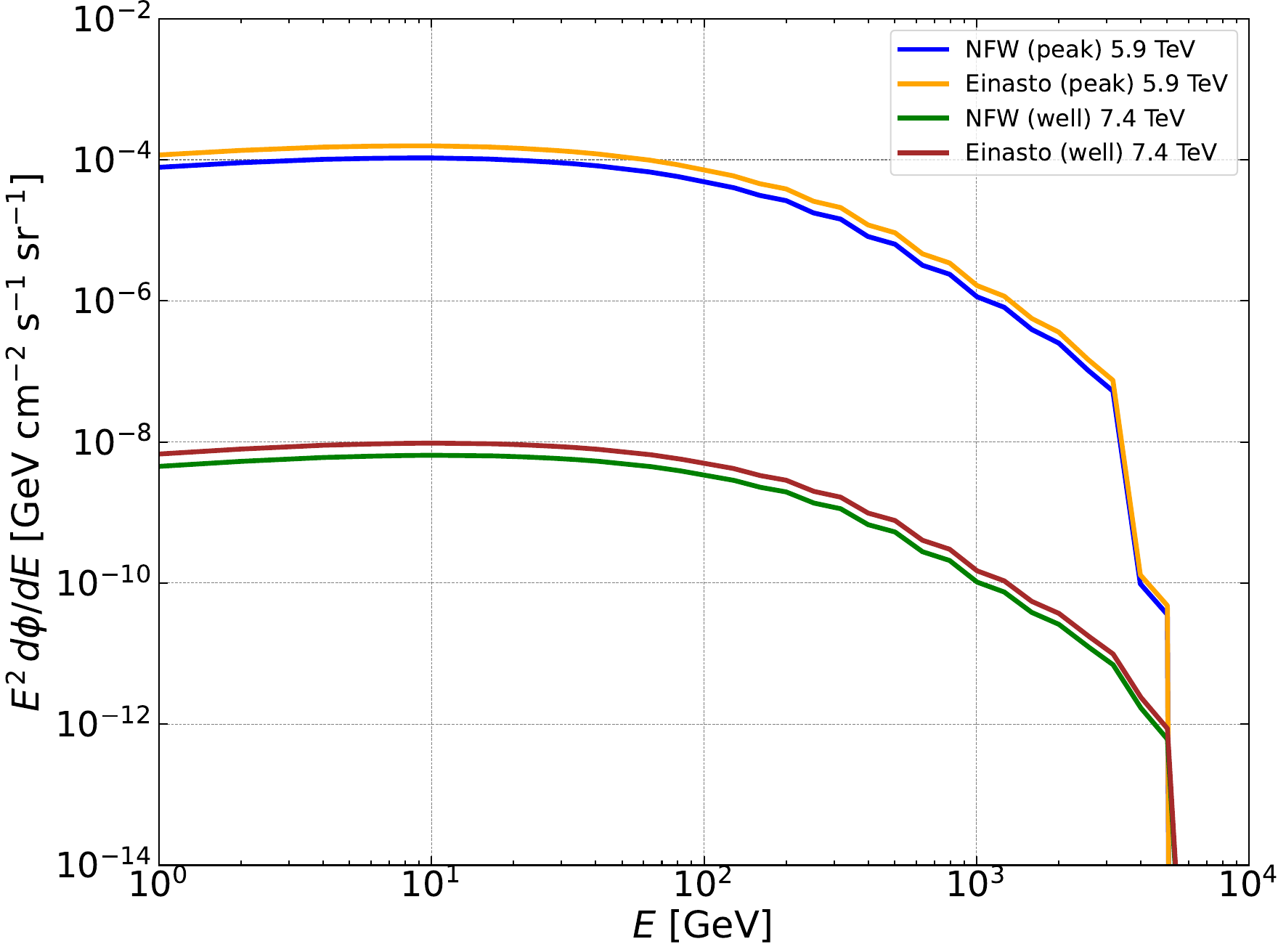}}\\\subfloat[total differential flux annihilation two photons]{
   \label{fluxtotkappa0}
    \includegraphics[width=0.5\textwidth]{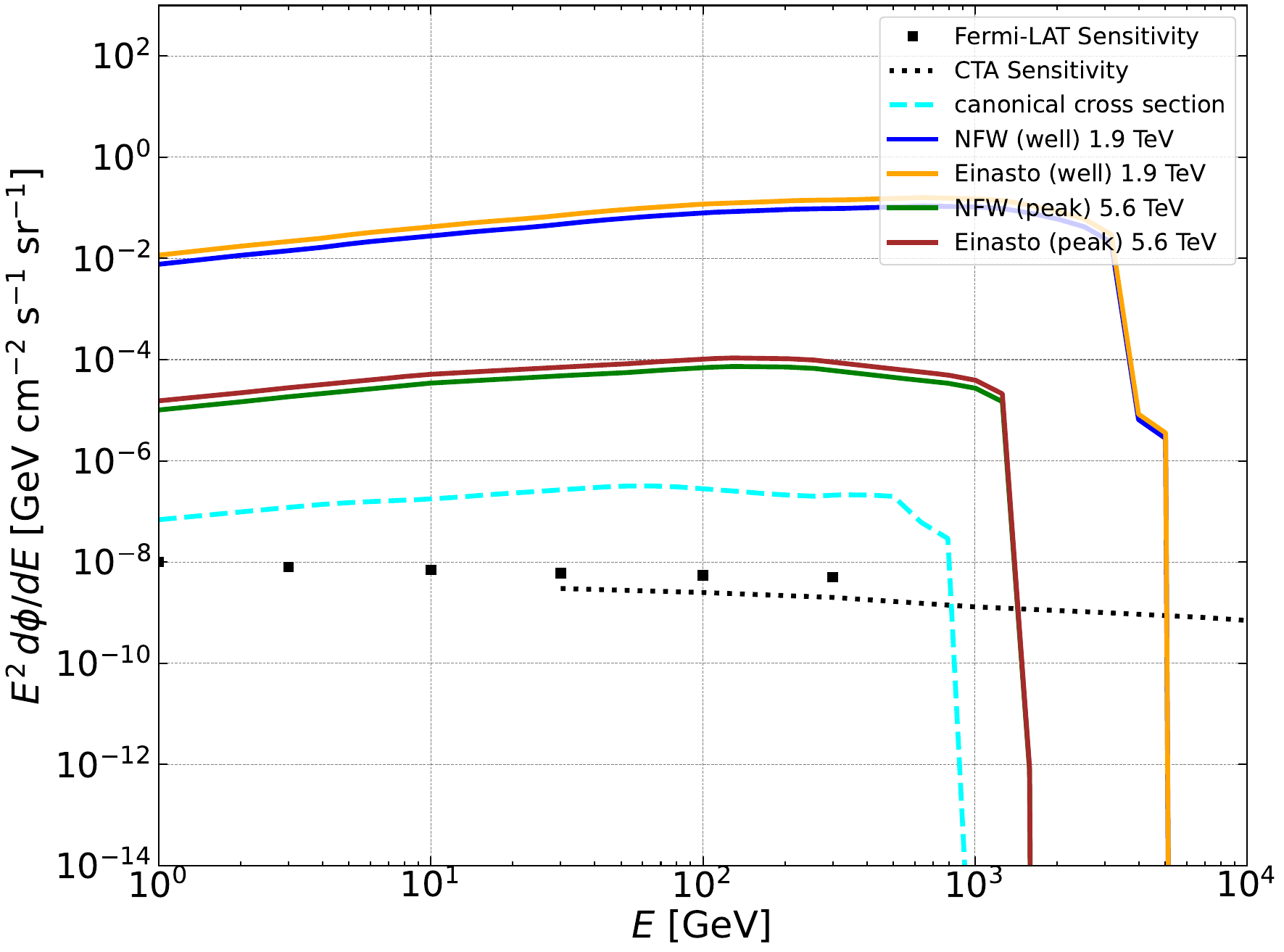}}
    \caption{Differential photon flux from DM particle annihilation for various final states. The solid lines represent the predictions of our model for different DM masses, assuming Einasto and NFW density profiles, with $\lambda=5$ and $\kappa=0$. The cyan dashed line corresponds to the flux associated with the canonical thermal annihilation cross section. The black points and the black dotted line represent the differential sensitivities of the Fermi-LAT and CTA telescopes, respectively, toward the Galactic Center.}
 \label{Fluxkappa}
\end{figure}

\begin{figure}
 \centering
  \subfloat[$v_{0}a_{0}\longrightarrow\gamma\gamma$]{
   \label{flux0}
\includegraphics[width=0.5\textwidth]{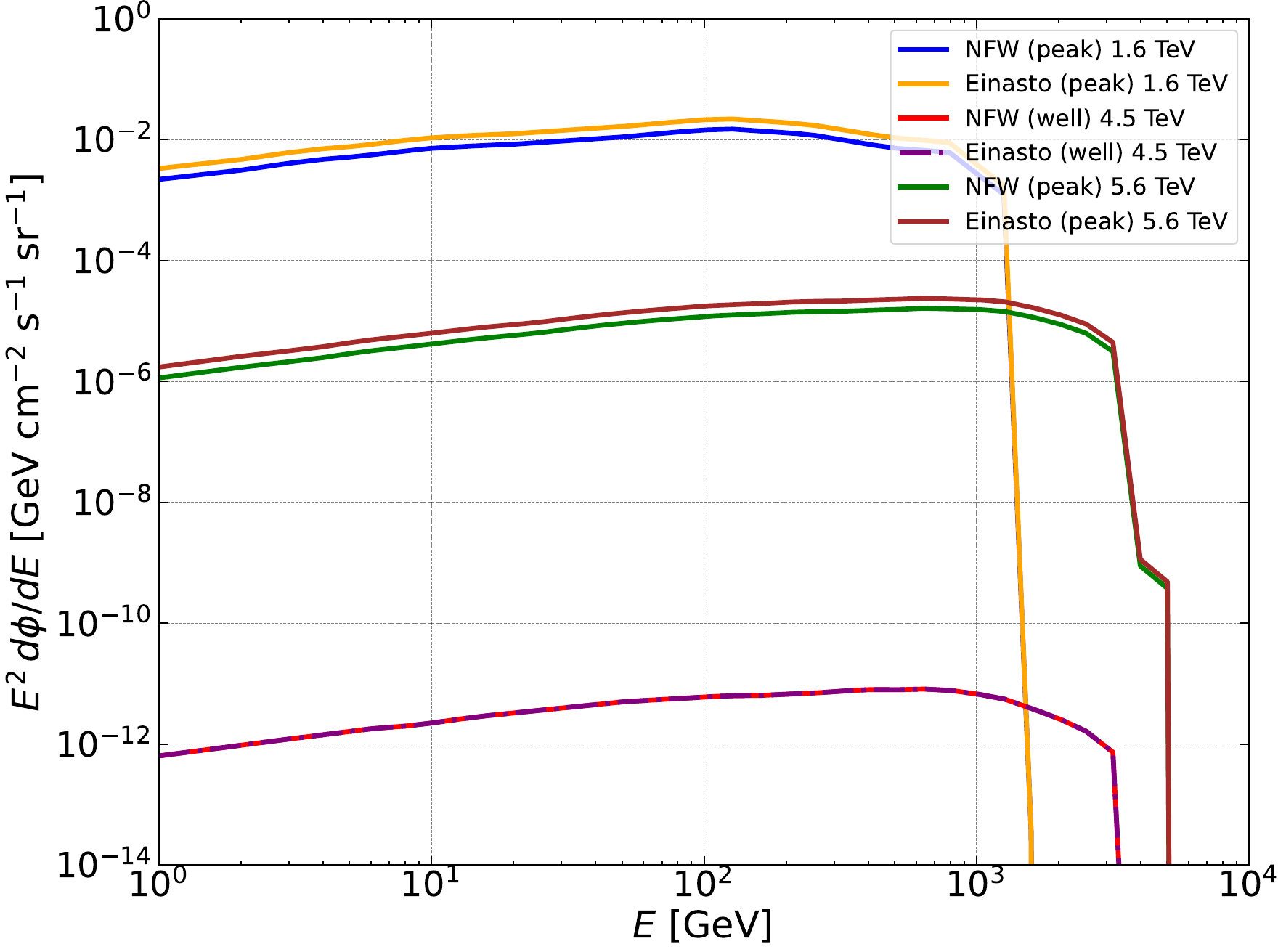}}
  \subfloat[$v_{0}a_{0}\longrightarrow W^{+}W^{-}$]{
   \label{flux0.5}
    \includegraphics[width=0.5\textwidth]{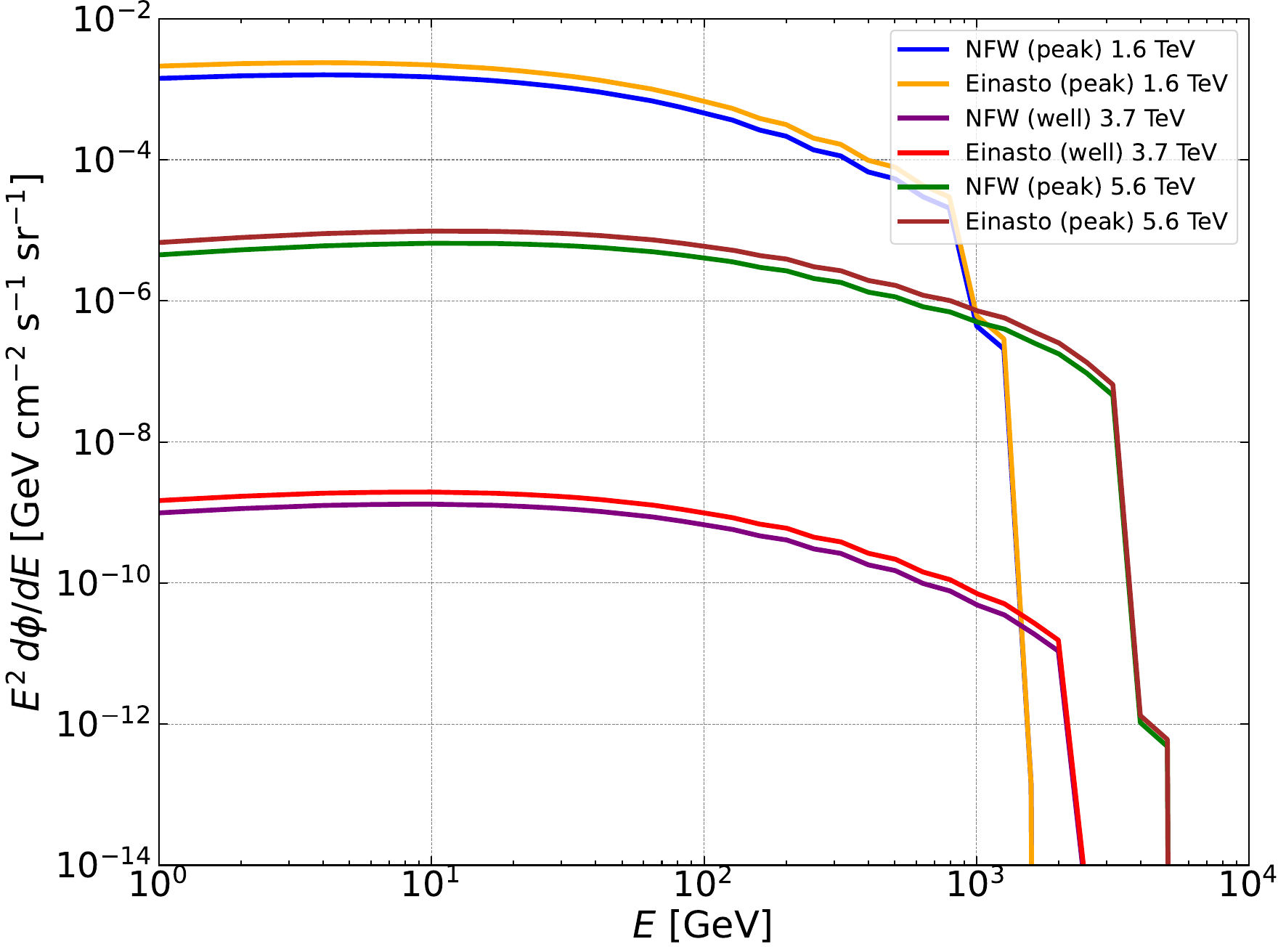}}\\
    \subfloat[$v_{0}a_{0}\longrightarrow ZZ$]{
   \label{flux2}
    \includegraphics[width=0.5\textwidth]{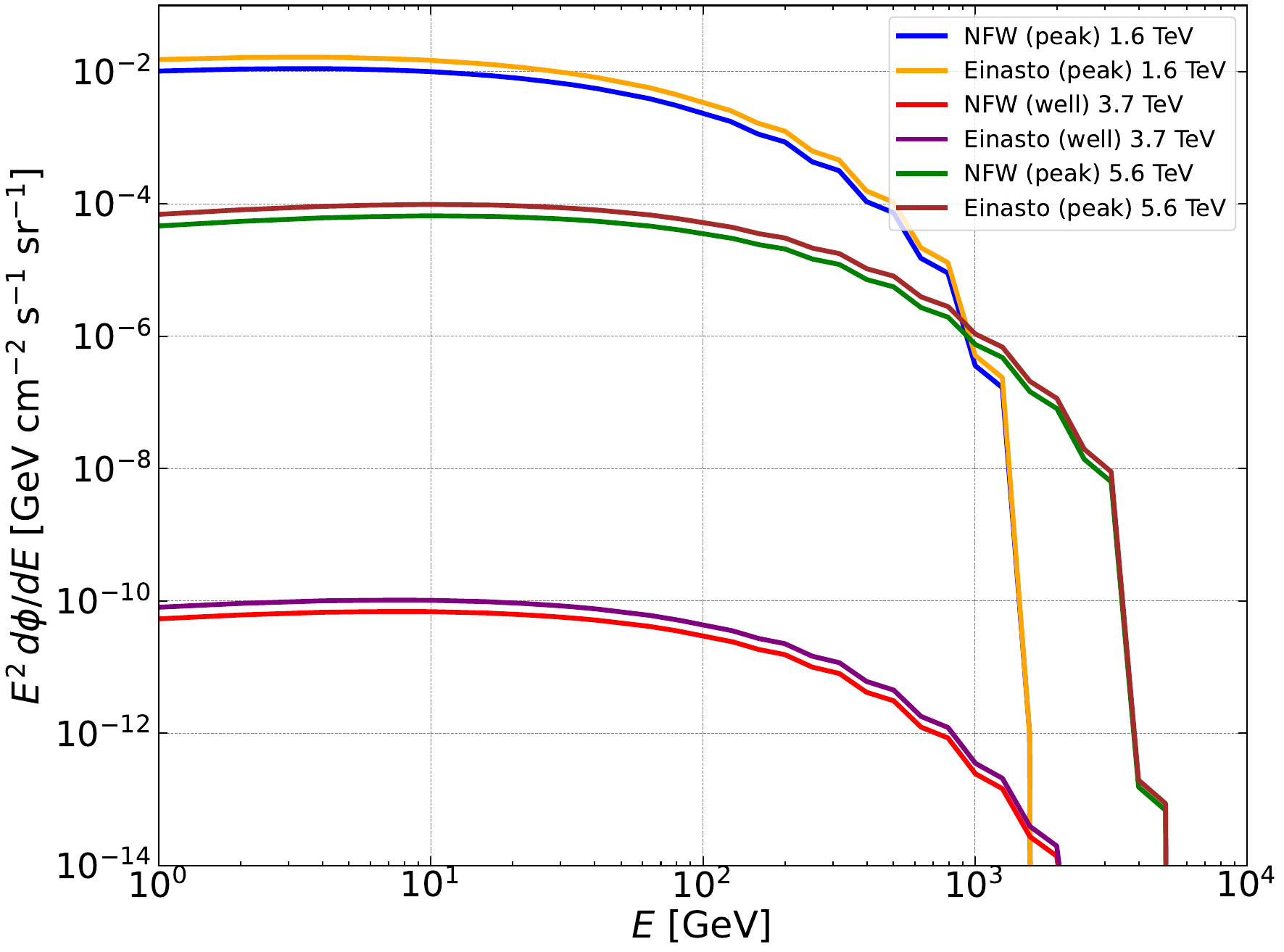}}
  \subfloat[$v_{0}a_{0}\longrightarrow hh$]{
   \label{flux4pi}
    \includegraphics[width=0.5\textwidth]{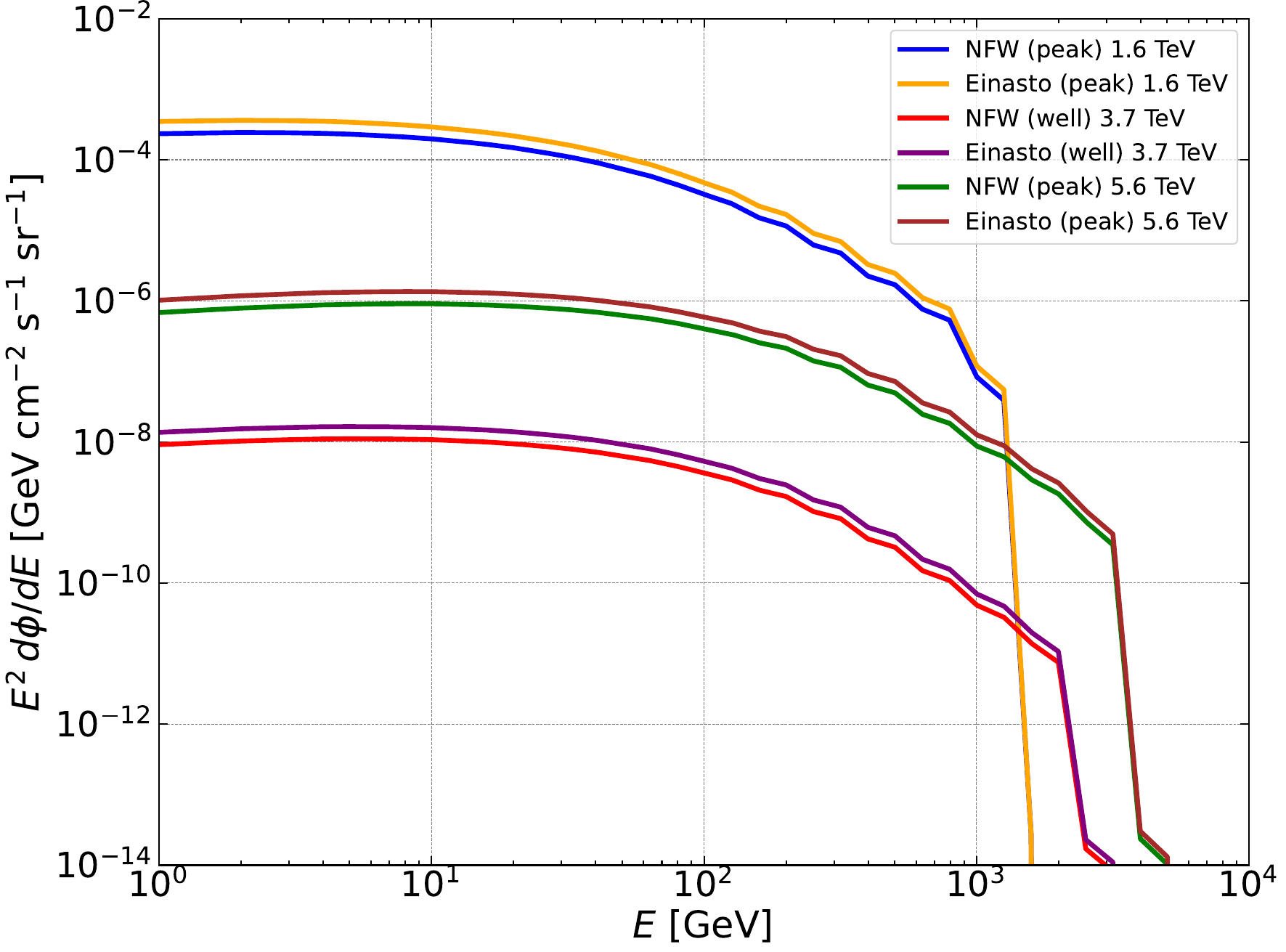}}\\
    \subfloat[total differential flux annihilation two photons ]{
   \label{fluxtotalkappa1}
    \includegraphics[width=0.5\textwidth]{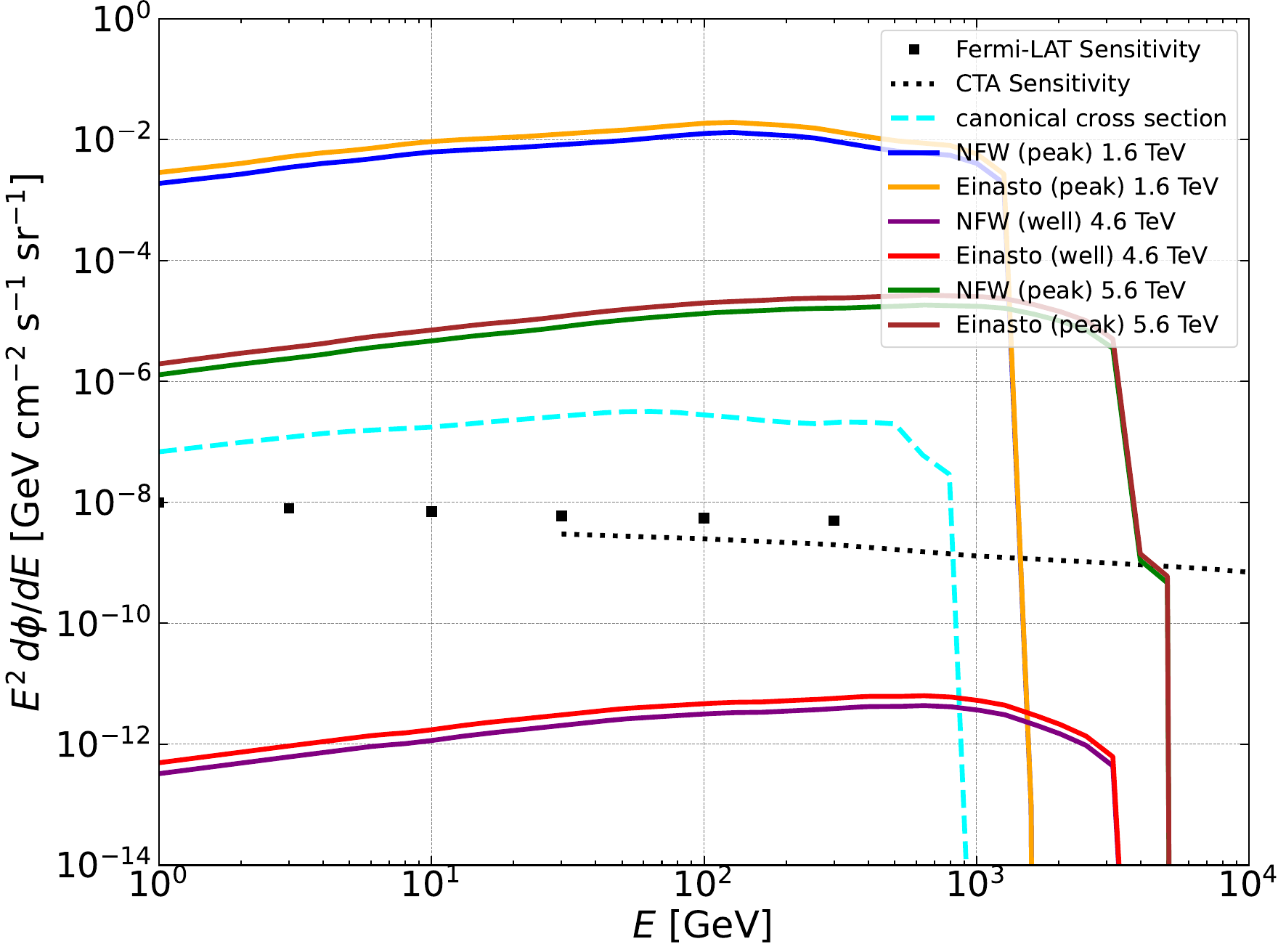}}
 \caption{Differential photon flux from DM particle annihilation for various final states. The solid lines represent the predictions of our model for different DM masses, assuming Einasto and NFW density profiles, with $\lambda=5$ and $\kappa=1$. The cyan dashed line corresponds to the flux associated with the canonical thermal annihilation cross section. The black points and the black dotted line represent the differential sensitivities of the Fermi-LAT and CTA telescopes, respectively, toward the Galactic Center.}
 \label{Flux}
\end{figure}

Our results are shown in Figures \ref{Fluxkappa} and \ref{Flux}, for different final states ($\gamma \gamma$, $W^{+}W^{-}, ZZ, hh$), as well as for the total flux. In particular, we compute the photon flux for two extreme scenarios: when the DM mass corresponds to a Sommerfeld resonance and when it lies at a local minimum of the Sommerfeld enhancement factor.

In Figures \ref{fluxtotkappa0} and \ref{fluxtotalkappa1}, we compare the total flux predicted by our model with the differential sensitivities of the Fermi-LAT~\cite{Ibarra_2012} and CTA~\cite{Silverwood_2015} in the Galactic Center region. The Fermi-LAT limits are based on several years of accumulated data in the energy range $5–300$ GeV, while the CTA projections assume 100 hours of observation in the energy range between $25$ GeV and $10$ TeV. In the case of Figure \ref{fluxtotkappa0}, we observe that the predictions of our model lie well above the experimental sensitivities, and are therefore excluded for the considered mass values, regardless of whether they correspond to maxima or minima of the Sommerfeld enhancement factor.


On the other hand, the scenario shown in Figure \ref{fluxtotalkappa1} exhibits a different behavior. Independently of the adopted DM profile, the resonance corresponding to $m_V=1.6$ TeV is excluded. However, for $m_V=5.6$ TeV, the resonance displays a narrow energy region, around $E \approx  4$ TeV, in which the model prediction approaches the sensitivity threshold of CTA. In this range, the flux crosses this limit, giving rise to a region that is partially compatible with current constraints, thereby opening the possibility of probing this region of parameter space with future observations. Finally, the local minimum at $4.6$ TeV lies below the limits of both experiments. Nevertheless, since the predicted flux is several orders of magnitude below the sensitivities of both telescopes, the signal may be strongly suppressed, making its detection with current instruments challenging.

\newpage
\section{Summary}\label{Summary}
In this work, we have studied the Sommerfeld enhancement in a model where the DM candidate is the neutral component of a new massive vector field in the fundamental representation of $SU(2)_{L}$. We computed the annihilation cross section including the Sommerfeld enhancement factor and analyzed its implications for indirect detection. Our results show that SE plays a key role in the non-relativistic regime characteristic of DM halos.

We find the appearance of resonances in the annihilation cross section for masses in the $2$–$10$ TeV range, capable of enhancing it by several orders of magnitude and more pronounced than those in other models, such as the spin-1/2 doublet and the spin-1 triplet. These resonances give rise to regions of parameter space that are excluded by $\gamma$-ray line searches, as well as to modifications in the predicted gamma-ray flux. Their structure depends on the Higgs portal coupling $\lambda$, while non-minimal interactions ($\kappa\neq0$) introduce additional resonant features without significantly altering their overall amplitude.

Regarding the $\gamma$-ray flux, the predictions depend critically on the proximity to these resonances. In the resonant case, the flux can exceed current experimental sensitivities, leading to excluded regions in agreement with $\gamma$-ray line searches, whereas in suppressed regions the signal remains undetectable. However, in intermediate scenarios, the predicted flux lies close to the projected sensitivity of CTA.

These results highlight the importance of accounting for non-perturbative effects in DM phenomenology, particularly in the non-relativistic regime, and demonstrate the potential of indirect detection to probe and constrain such scenarios.

\section*{Acknowledgments}
This work was partially found by Fondecyt grant 1230110, grant PIIC 2021-II UTFSM, ANID BECA/DOCTORADO NACIONAL 21210616, ANID PIA/APOYO AFB220004, and ANID—The Millennium Science Initiative Program ICN2019\_044. We are very thankful to Motoko Fujiwara and Patricio Escalona for discussions and suggestions that contributed significantly to improve the quality of this paper.


\newpage
\appendix

\section{Electroweak Matrix}\label{A}
\subsection{Gauge-Dark Matter}
In the context of non-relativistic DM, we need to determine the potentials that interact with these particles in a low-velocity regime. For this, we need to determine the real potential (cubic terms) of the Lagrangian~\eqref{Lag1} or effective action of $SU(2)_L$, which is the electroweak matrix, where the conserved currents interact with the gauge fields of the SM, resulting in the following expression:

\begin{align}\label{effeaction1}
S_{eff,1}\approx\int d^{4}x~ \left[J_{\mu}^{W^{+}}(x)W^{+\mu}(x)+J_{\mu}^{W^{-}}(x)W^{-\mu}(x)+J_{\mu}^{Z}(x)Z^{\mu}(x)+J_{\mu}^{A}(x)A^{\mu}(x)\right].
\end{align}

After integrating out the gauge bosons in the path integral, we obtain the effective action dependent on the currents. We need to calculate the conserved currents from the cubic terms of the Lagrangian \eqref{Lag1}; the details of the calculation are in ~\cite{Chowdhury_2017}.

\begin{align}\label{acceffnon1}
S_{eff,1}=-\int\dfrac{d^{4}xd^{3}y}{8\pi|\vec{x}-\vec{y}|}(&J_{A}^{0}(x)J_{A}^{0}(x^{0},\vec{y})+J_{Z}^{0}(x)J_{Z}^{0}(x^{0},\vec{y})e^{-M_{Z}|\vec{x}-\vec{y}|}\nonumber \\ &+2J_{W^{+}}^{0}(x)J_{W^{-}}^{0}(x^{0},\vec{y})e^{-M_{W}|\vec{x}-\vec{y}|}).
\end{align}


It is essential to consider only pairs in the form of vectors, as indicated in \eqref{vectornonrel}, while other particle combinations are prohibited. Consequently, the currents in the non-relativistic limit are as follows:


 \begin{align}\label{JA}
J_{A}^{0}(x)J_{A}^{0}(x^{0},\vec{y})=-2g^{2}(c_{W}(1-(\kappa_{1}+\kappa{2})-s_{W}t_{W}^{2})^{2}&(-[v_{+}^{\dagger}(x)v_{-}^{\dagger}(x^{0},\vec{y})v_{+}(x)v_{-}(x^{0},\vec{y})],
\end{align}

\begin{align}\label{JZ}
J_{Z}^{0}(x)J_{Z}^{0}(x^{0},\vec{y})&=-g^{2}(c_{W}(\kappa_{1}+\kappa_{2})+s_{W})^{2}[v_{+}^{\dagger}(x)v_{-}^{\dagger}(x^{0},\vec{y})v_{+}(x)v_{-}(x^{0},\vec{y})\nonumber\\
&+\left(\dfrac{v_{0}^{\dagger}(x)v_{0}^{\dagger}(x^{0},\vec{y})}{\sqrt{2}}\dfrac{v_{0}(x)v_{0}(x^{0},\vec{y})}{\sqrt{2}} \right)\nonumber\\
&+\left(\dfrac{a_{0}^{\dagger}(x)a_{0}^{\dagger}(x^{0},\vec{y})}{\sqrt{2}}\dfrac{a_{0}(x)a_{0}(x^{0},\vec{y})}{\sqrt{2}} \right)\nonumber\\
&-\left(v_{0}^{\dagger}(x)a_{0}^{\dagger}(x^{0},\vec{y})v_{0}(x)a_{0}(x^{0},\vec{y}) \right)],
\end{align}

\begin{align}\label{JW}
J_{W_{+}}^{0}(x)J_{W_{-}}^{0}(x^{0},\vec{y})=&-\left[\left(\dfrac{3}{8}-2\sqrt{2}\right)(\kappa_{1}+\kappa_{2})+2\sqrt{2}\right]^{2}\nonumber\\ &g^{2}\left[v_{0}^{\dagger}(x)a_{0}^{\dagger}(x^{0},\vec{y})v_{+}(x)v_{-}(x^{0},\vec{y})-v_{-}^{\dagger}(x)v_{+}^{\dagger}(x^{0},\vec{y})a_{0}(x)v_{0}(x^{0},\vec{y}) \right],
\end{align}

where $s_{W}=\sin\theta_{W}$, $c_{W}=\cos\theta_{W}$ and $t_{W}=\tan\theta_{W}$. Here the terms that are omitted are the ones that do not lead to pairs $v_{0}v_{0}$, $a_{0}a_{0}$, $v_{0}a_{0}$ and $v_{+}v_{-}$. By plugging these currents in equation \eqref{acceffnon1}, we obtain:

\begin{align}\label{acNR}
S_{eff,1_{NR}}=-\int d^{4}xd^{3}y~s(x,\vec{y})^{\dagger}V_{g-V}(|\vec{x}-\vec{y}|)s(x,\vec{y}),
\end{align} 

where $V_{g-V}(|\vec{x}-\vec{y}|)$ is the electroweak matrix, with $r=|\vec{x}-\vec{y}|$, therefore:

\begin{align}\label{ewpot}
V_{g-V}(r)=-\dfrac{g^{2}}{8\pi r} \begin{pmatrix}4a^{2}e^{-M_{Z}r} & 0 & 0 &0\\
0 & 4a^{2}e^{-M_{Z}r} & 0 & 0\\
0 & 0 & 16a^{2}e^{-M_{Z}r} & b^{2}e^{-M_{W}r}\\
0& 0 & b^{2}e^{-M_{W}r} &a^{2}e^{-M_{Z}r}-2(c_{W}(1-(\kappa_{1}+\kappa{2})-s_{W}t_{W}^{2})^{2} 
\end{pmatrix},
\end{align}
with $a=(c_{W}(\kappa_{1}+\kappa_{2})+s_{W})$ and $b=(\frac{3}{8}-2\sqrt{2})(\kappa_{1}+\kappa_{2})+2\sqrt{2}$.

\subsection{Higgs Boson-Dark Matter}
We are now employing the Feynman gauge. The masses of the Goldstone bosons, namely $\phi^{0}$, $\phi^{0\ast}$, and $\phi^{\pm}$, originating from the doublet \eqref{gold}, are $M_{Z}$ and $M_{W}$. The Higgs particle is significantly lighter than DM particles. The interacting Lagrangian \eqref{Lag1} between the Higgs Boson and the DM vector is as follows:

\begin{align}\label{lag H-V}
\mathcal{L}_{h-V}=-\lambda_{2}(\phi^{\dagger}\phi)(V_{\mu}^{\dagger}V^{\mu})-\lambda_{3}(\phi^{\dagger}V_{\mu})(V^{\mu\dagger}\phi)-\dfrac{\lambda_{4}}{2}\left[(\phi^{\dagger}V_{\mu})(\phi^{\dagger}V^{\mu})+(V^{\mu\dagger}\phi)(V_{\mu}^{\dagger}\phi)\right].    
\end{align}

In order to construct pairs such as $v_{0}v_{0}, v_{0}a_{0}, a_{0}a_{0}$, and $v_{+}v_{-}$, we need to substitute equations \eqref{vector} and \eqref{gold} into \eqref{lag H-V}. Consequently, we obtain:

\begin{align}\label{V h-V1}
V_{h-V}(r)=\dfrac{v^{2}}{32\pi M_V^{2} r}\begin{pmatrix}
8\lambda_{4}^{2}a & 0 & 0 & 0\\
0 & 8\lambda_{4}^{2}a & 0 & 0\\
0 & 0 & \dfrac{1}{2} (\lambda_{2}+\lambda_{3})^{2}a & (\lambda_{3}+2\lambda_{4})^{2}e^{-M_{W}r}\\
0 & 0 & (\lambda_{3}+2\lambda_{4})^{2}e^{-M_{W}r} & \lambda_{2}^{2}a
\end{pmatrix},
\end{align}

with $a=(2e^{-M_{h}r}+e^{-M_{Z}r})$,
this is electroweak matrix (Higgs Boson-DM interaction).

\section{Annihilation Matrix}\label{B}
\subsection{Gauge-Dark Matter Interaction}

Non-relativistic vector DM particles have the potential to annihilate into relativistic particles, specifically gauge bosons of $SU(2)_{L}$. In quantum mechanics (Schrödinger equation), these annihilations are represented by an imaginary potential or an absorptive term (quartic terms) in the action, which explains the disappearance of pairs of non-relativistic particles.

These absorptive terms originate from $S_{eff,2}$. To provide an interpretation, we can visualize $S_{eff,2}$ in terms of Feynman diagrams. In this representation, the matrix $\Delta$ characterizes the propagators of the gauge fields, and each $\mathcal{U}$ represents a vertex featuring two gauge fields and two vector DM fields, as described by Chowdhury et al.~\citep{Chowdhury_2017,GarciaCely:2014jha}.



Then, the effective action of absorptive terms is:

\begin{align}\label{seff2}
S_{eff,2}&=-i Tr\log\left(\Delta^{-1}\mathcal{U}\Delta^{-1}\mathcal{U}\right)\nonumber\\
&=\frac{i}{2\pi}\int d^{4}x~ Tr\left[\begin{pmatrix}
U_{AA}(x)&U_{AZ}(x)&U_{AW}(x)&U_{AW}^{\ast}(x)\\
U_{AZ}(x)&U_{ZZ}(x)&U_{ZW}(x)&U_{ZW}^{\ast}(x)\\
U_{AW}^{\ast}(x)&U_{ZW}^{\ast}(x)&U_{WW}(x)&0\\
U_{AW}^{\ast}(x)&U_{ZW}^{\ast}(x)&0&U_{WW}^{\ast}(x)\\
\end{pmatrix}\right]^{2}.
\end{align}








We need only pairs like $V_{0}V_{0}, V_{0}^{\ast}V_{0}^{\ast}, V_{0}V_{0}^{\ast}$ and $V_{+}V_{-}$, then

\begin{align}\label{sQs1}
S_{eff,2}&=\dfrac{i}{2\pi}\int~d^{4}x\left(U_{AA}(x)^{2}+U_{ZZ}(x)^{2}+2U_{WW}(x)^{2}+2U_{AZ}(x)^{2} \right)\nonumber\\ &=-\int~d^{4}x \dfrac{g^{2}}{8}[2W_{\mu}^{+}V_{\nu}^{-}V^{+~\nu}W^{-~\mu}+2W_{\mu}^{-}(V_{\nu}^{0})^{\ast}V^{0~\nu}W^{+~\mu}\nonumber\\
&+A_{\mu}\left\lbrace \sin^{2}\theta_{W}(1-\tan\theta_{W})(V_{\nu}^{0})^{\ast}V^{0~\nu}+2\sin^{2}\theta_{W}V_{\nu}^{-}V^{+~\nu} \right\rbrace A^{\mu}\nonumber\\
&+Z_{\mu}\left\lbrace (\tan^{2}\theta_{W}\sin^{2}\theta_{W}-\sin\theta_{W}\cos\theta_{W})(V_{\nu}^{0})^{\ast}V^{0~\nu}+2\sin^{2}\theta_{W}V_{\nu}^{-}V^{+~\nu} \right\rbrace Z^{\mu}\nonumber\\
&+A_{\mu}( (\sin\theta_{W}(2\cos\theta_{W}-\sin\theta_{W} (1+\tan\theta_{W} ))(V_{\nu}^{0})^{\ast}V^{0~\nu}\nonumber\\
&+\cos (2\theta_{W})\tan\theta_{W}V_{\nu}^{-}V^{+~\nu}) Z^{\mu}\nonumber\\
&+Z_{\mu}( \sin\theta_{W}(\cos\theta_{W}-\sin\theta_{W})(V_{\nu}^{0})^{\ast}V^{0~\nu}\nonumber\\
&+\tan\theta_{W}(\cos^{2}\theta_{W}-\sin^{2}\theta_{W})V_{\nu}^{-}V^{+~\nu})  A^{\mu}](1+1+1+1)\nonumber\\
&=-\dfrac{g^{2}}{2}[2W_{\mu}^{+}V_{\nu}^{-}V^{+~\nu}W^{+~\mu}+2W_{\mu}^{-}(V_{\nu}^{0})^{\ast}V^{0~\nu}W^{+~\mu}\nonumber\\
&+A_{\mu}\left\lbrace \sin^{2}\theta_{W}(1-\tan\theta_{W})(V_{\nu}^{0})^{\ast}V^{0~\nu}+2\sin^{2}\theta_{W}V_{\nu}^{-}V^{+~\nu} \right\rbrace A^{\mu}\nonumber\\
&+Z_{\mu}\left\lbrace (\tan^{2}\theta_{W}\sin^{2}\theta_{W}-\sin\theta_{W}\cos\theta_{W})(V_{\nu}^{0})^{\ast}V^{0~\nu}+2\sin^{2}\theta_{W}V_{\nu}^{-}V^{+~\nu} \right\rbrace Z^{\mu}\nonumber\\
&+A_{\mu}( (\sin\theta_{W}(2\cos\theta_{W}-\sin\theta_{W} (1+\tan\theta_{W} ))(V_{\nu}^{0})^{\ast}V^{0~\nu}\nonumber\\
&+\cos (2\theta_{W})\tan\theta_{W}V_{\nu}^{-}V^{+~\nu}) Z^{\mu}\nonumber\\
&+Z_{\mu}( \sin\theta_{W}(\cos\theta_{W}-\sin\theta_{W})(V_{\nu}^{0})^{\ast}V^{0~\nu}\nonumber\\
&+\tan\theta_{W}(\cos^{2}\theta_{W}-\sin^{2}\theta_{W})V_{\nu}^{-}V^{+~\nu})  A^{\mu}].
\end{align}

In terms of s'vectors, the effective action is
\begin{align}
  S_{eff,2}=2i\int d^{4}x d^{3}y~s(x,\vec{y})^{\dagger}\Gamma_{gauge}\delta^{3}(|x-y|)s(x,\vec{y}).
\end{align}

The Q's terms are:

\begin{align}\label{Qww}
U_{WW}^{2}(x)&=\dfrac{g^{4}}{16}(V_{\nu}^{-}V^{+~\nu}+(V_{\nu}^{0})^{\ast}V^{0~\nu})^{2}\\
&=\dfrac{g^{4}}{16}(V_{i}^{-}V^{+~j}V_{i}^{-}V^{+~j}+(V_{i}^{0})^{\ast}V^{0~j}(V_{i}^{0})^{\ast}V^{0~j}\nonumber\\
&+2(V_{i}^{0})^{\ast}V^{0~j}(V_{i}^{-})^{\ast}V^{+~j}),\nonumber
\end{align}

\begin{align}\label{Qaa}
U_{AA}^{2}(x)=\dfrac{e^{4}}{64}\left[(1-\tan\theta_{W})^{2}((V_{\nu}^{0})^{\ast}V^{0~\nu})^{2}+4(V_{\nu}^{-}V^{+~\nu})^{2}+4(1-\tan\theta_{W})(V_{\nu}^{0})^{\ast}V^{0~\nu}V_{\nu}^{-}V^{+~\nu}  \right],    
\end{align}

\begin{align}\label{Qzz}
U_{ZZ}^{2}(x)=\dfrac{g^{4}}{64}&[(\tan^{2}\theta_{W}\sin^{2}\theta_{W}-\sin\theta_{W}\cos\theta_{W})^{2}((V_{\nu}^{0})^{\ast}V^{0~\nu})^{2}+4\sin^{4}\theta_{W}(V_{\nu}^{-}V^{+~\nu})^{2}\nonumber\\     &-4\sin^{2}\theta_{W}(\tan^{2}\theta_{W}\sin^{2}\theta_{W}-\sin\theta_{W}\cos\theta_{W})^{2}(V_{\nu}^{0})^{\ast}V^{0~\nu}V_{\nu}^{-}V^{+~\nu} ],
\end{align}

\begin{align}\label{Qaz2}
U_{AZ}^{2}(x)&=\dfrac{e^{2}g^{2}}{64}[(\cos\theta_{W}-\sin\theta_{W})^{2} ((V_{\nu}^{0})^{\ast}V^{0~\nu})^{2}+\dfrac{\cos^{2}(2\theta_{W})}{\cos^{2}\theta_{W}}{\cos\theta_{W}}(V_{\nu}^{-}V^{+~\nu})^{2}\nonumber\\
&+(\cos\theta_{W}-\sin\theta_{W})^{4}(1+\tan\theta_{W})^{2}(V_{\nu}^{0})^{\ast}V^{0~\nu}V_{\nu}^{-}V^{+~\nu}].
\end{align}

However, we need to establish a relationship between polarization vectors and angular momentum, as discussed in \citep{Abe_2021}, where:

\begin{align}\label{Smatrix}
\epsilon_{i}(p_{-})\epsilon_{j}(p_{+})\longrightarrow S_{ij}^{J,J_{z}}.    
\end{align}

The annihilation matrices (Gauge-DM) are:

\begin{align}\label{MatrixW}
\Gamma_{WW}^{J=0}=\dfrac{2g^{4}}{64\sqrt{3}\pi M_V^{2}}\begin{pmatrix}
2 & 0 & 0 & 0\\
0 & 2 & 0 & 0\\
0 & 0 & 1 & 1\\
0 & 0 & 1 & 1
\end{pmatrix},
\end{align}

\begin{align}
    \Gamma_{WW}^{J=2}=\dfrac{g^{4}}{64\pi M_V^{2}}\left(\dfrac{1+3\cos(2\theta)}{2\sqrt{6}}+\sin^{2}\theta \right)\begin{pmatrix}
2 & 0 & 0 & 0\\
0 & 2 & 0 & 0\\
0 & 0 & 1 & 1\\
0 & 0 & 1 & 1
\end{pmatrix},
\end{align}

\begin{align}\label{MatrixA}
\Gamma_{AA}^{J=0}=\dfrac{e^{4}}{128\sqrt{3}\pi M_V^{2}}\begin{pmatrix}
2a^{2} & 0 & 0 & 0\\
0 & 2a^{2} & 0 & 0\\
0 & 0 & a^{2} & 4a^{2}\\
0 & 0 & 4a^{2} & 4
\end{pmatrix},
\end{align}

\begin{align}
    \Gamma_{AA}^{J=2}=\dfrac{e^{4}}{256\pi M_V^{2}}\left(\dfrac{1+3\cos(2\theta)}{2\sqrt{6}}+\sin^{2}\theta \right)\begin{pmatrix}
2a^{2} & 0 & 0 & 0\\
0 & 2a^{2} & 0 & 0\\
0 & 0 & a^{2} & 4a^{2}\\
0 & 0 & 4a^{2} & 4
\end{pmatrix},
\end{align}
with $a=(1-\tan\theta_{W})$.

\begin{align}\label{MatrixZ}
\Gamma_{ZZ}^{J=0}&=\dfrac{g^{4}}{128\sqrt{3}\pi M_V^{2}}\begin{pmatrix}
2b^{2} & 0 & 0 & 0\\
0 & 2b^{2} & 0 & 0\\
0 & 0 & 2b^{2} & -4c\\
0 & 0 & -4c & 4\sin^{4}\theta_{W}
\end{pmatrix},\\\Gamma_{ZZ}^{J=2}&=\dfrac{g^{4}}{256\pi M_V^{2}}\left(\dfrac{1+3\cos(2\theta)}{2\sqrt{6}}+\sin^{2}\theta \right)\begin{pmatrix}
2b^{2} & 0 & 0 & 0\\
0 & 2b^{2} & 0 & 0\\
0 & 0 & 2b^{2} & -4c\\
0 & 0 & -4c & 4\sin^{4}\theta_{W}
\end{pmatrix},\nonumber
\end{align}

with $b=(\tan^{2}\theta_{W}\sin^{2}\theta_{W}-\sin\theta_{W}\cos\theta_{W})$ and $c=\sin^{2}\theta_{W}\tan\theta_{W}(\sin^{2}\theta_{W}-\cos^{2}\theta_{W})$.

\begin{align}\label{MatrixAZ}
\Gamma_{AZ}^{J=0}&=\dfrac{e^{2}g^{2}}{128\sqrt{3}\pi M_V^{2}}\begin{pmatrix}
d^{2} & 0 & 0 & 0\\
0 & d^{2} & 0 & 0\\
0 & 0 & d^{2} & f\\
0 & 0 & f & \dfrac{\sin^{2}(2\theta_{W})}{\cos^{2}\theta_{W}}
\end{pmatrix},\\\Gamma_{AZ}^{J=2}&=\dfrac{e^{2}g^{2}}{256\pi M_V^{2}}\left(\dfrac{1+3\cos(2\theta)}{2\sqrt{6}}+\sin^{2}\theta \right)\begin{pmatrix}
d^{2} & 0 & 0 & 0\\
0 & d^{2} & 0 & 0\\
0 & 0 & d^{2} & f\\
0 & 0 & f & \dfrac{\sin^{2}(2\theta_{W})}{\cos^{2}\theta_{W}}
\end{pmatrix},\nonumber
\end{align}

with $d=(\cos\theta_{W}-\sin\theta_{W})$ and $f=(\cos\theta_{W}-\sin\theta_{W})^{4}(1+\tan\theta_{W})^{2}$.

\subsection{Higgs-Dark Matter Interaction}

From the Lagrangian \eqref{lag H-V}, we can extract the quartic terms in the effective action. These terms provide the annihilation matrix for the interaction between the Higgs boson and the DM vector. We need to operate on the vector in $SU(2)_{L}$ \eqref{vector} and the Higgs boson doublet \eqref{gold}. Consequently,

\begin{align}\label{cuartico h-V}
\mathcal{L}_{Quartic~ h-V}&=-\lambda_{2}[\left(\phi^{-}\phi^{+}+\dfrac{h^{2}}{2}-i\dfrac{(\phi^{0\ast}h+h^{\ast}\phi^{0})}{\sqrt{2}}-(\phi^{0})^{2} \right)V_{\mu}^{-}V^{+\mu}\\
&+\left(\phi^{-}\phi^{+}+\dfrac{h^{2}}{2}-i\dfrac{(\phi^{0\ast}h+h^{\ast}\phi^{0})}{\sqrt{2}}-(\phi^{0})^{2} \right)V_{\mu}^{0\ast}V^{0\mu}]\nonumber\\
&-\lambda_{3}[\phi^{-}\phi^{+}V_{\mu}^{+}V^{-\mu}+\dfrac{1}{\sqrt{2}}(i\phi^{0}+h)\phi^{-}V_{\mu}^{+}V^{0\mu\ast}+\dfrac{1}{\sqrt{2}}(-i\phi^{0\ast}+h^{\ast})\phi^{+}V_{\mu}^{0}V^{-\mu}\nonumber\\
&+\left(\dfrac{h^{2}}{\sqrt{2}}-(\phi^{0})^{2} \right)V_{\mu}^{0}V^{0\mu\ast}]\nonumber\\
&+\left(\dfrac{h^{2}}{\sqrt{2}}-(\phi^{0})^{2} \right)V_{\mu}^{0}V^{0\mu\ast}]\nonumber\\
&-\lambda_{4}[\left(-i\phi^{0\ast}+\dfrac{h^{\ast}}{\sqrt{2}}\right)\phi^{-}V^{\mu +}V^{0\mu}+\left(-i\phi^{0\ast}+\dfrac{h^{\ast}}{\sqrt{2}} \right)\phi^{-}V^{\mu 0}V^{+\mu}\nonumber\\
&+\left(\dfrac{h^{\ast}}{2}-i\sqrt{2}h^{\ast}\phi^{0\ast}+(\phi^{0\ast})^{2} \right)V_{\mu}^{0}V^{0\mu}+\left(i\phi^{0}+\dfrac{h}{\sqrt{2}} \right)\phi^{+}V^{-\mu}V_{\mu}^{0\ast}\nonumber\\
&+\left(i\phi^{0}+\dfrac{h}{\sqrt{2}} \right)\phi^{+}V^{0\mu\ast}V_{\mu}^{-}+\left(-(\phi^{0})^{2}+\dfrac{h^{2}}{2} \right)V^{0\mu\ast}V_{\mu}^{0\ast}].\nonumber
\end{align}

The permissible terms are $V_{0}V_{0}, V_{0}V_{0}^{\ast}, V_{0}^{\ast}V_{0}^{\ast}$, and $V_{+}V_{-}$, as we require neutral terms. The resulting annihilation matrix is as follows:

\begin{align}\label{Matrix Anni}
\Gamma_{h-V}=\dfrac{1}{32\pi M_V^{2}}\begin{pmatrix}
\left(\dfrac{3}{2}+\sqrt{2} \right)^{2}\lambda_{4}^{2} & 0 & 0 & 0\\
0 & \dfrac{9}{4}\lambda_{4}^{2} & 0 & 0\\
0 & 0 & \dfrac{1}{4}(7\lambda_{2}^{2}+6\lambda_{2}\lambda_{3}+5\lambda_{3}^{2}) & \dfrac{1}{2}(\lambda_{3}+2\lambda_{4})^{2}\\
0 & 0 & \dfrac{1}{2}(\lambda_{3}+2\lambda_{4})^{2} & \dfrac{3}{4}\lambda_{2}^{2}+(\lambda_{2}+\lambda_{3})^{2}
\end{pmatrix}.
\end{align}

\bibliographystyle{unsrt}
\bibliography{bib}
\end{document}